\documentclass[aps,pra,twocolumn,a4paper,reprint,superscriptaddress,floatfix]{revtex4-1} 
\pdfoutput=1
\usepackage[colorlinks=true,citecolor=blue,linkcolor=red,urlcolor=blue]{hyperref}
\usepackage{amsmath}
\usepackage{amssymb}
\usepackage{dsfont}
\usepackage{verbatim}
\usepackage{graphicx}
\usepackage{tabularx}
\usepackage{xcolor}
\usepackage{mathtools}
\usepackage{bbold}
\usepackage{blkarray}
\usepackage{appendix}
\usepackage[shortlabels]{enumitem}
\usepackage{latexsym}
\usepackage{array}
\usepackage{colortbl}
\usepackage{bm} 
\usepackage[T1]{fontenc}
\usepackage{lipsum}  
\usepackage[english]{babel}
\usepackage{upgreek}
\usepackage{physics}
\usepackage{bookmark}
\usepackage[normalem]{ulem}

\usepackage{thmtools}

\usepackage{accents}

\usepackage{dsfont} 
\usepackage{hyperref}
\usepackage{nicefrac}
\hypersetup{colorlinks=true,        
    linkcolor=red,         
    citecolor=blue,        
    filecolor=black,      
    urlcolor=blue }

\usepackage{bbm}
\newcommand{\mrm}{\mathrm}

\newcommand{\mc}[1]{\mathcal{#1}}
\newcommand{\bB}{\mathbf B}
\newcommand{\La}{\Lambda}
\newcommand{\id}{\mathbbm{1}}

\begin{document}

\title{Exchange-only qubit stabilized by a single-spin qubit}

\author{Irina Heinz}
\email{i.heinz@fz-juelich.de}
\affiliation{Institute for Theoretical Nanoelectronics (PGI-2), Forschungszentrum J\"ulich, D-52425 J\"ulich, Germany}
\affiliation{Institute for Quantum Information, RWTH Aachen University, D-52056 Aachen, Germany}

\author{Mira Sharma}
\email{mrsharma@physik.rwth-aachen.de}
\affiliation{Institute for Quantum Information, RWTH Aachen University, D-52056 Aachen, Germany}

\author{Joris Kattem\"{o}lle}
\affiliation{Institute for Theoretical Nanoelectronics (PGI-2), Forschungszentrum J\"ulich, D-52425 J\"ulich, Germany}
\affiliation{Institute for Quantum Information, RWTH Aachen University, D-52056 Aachen, Germany}

\begin{abstract}
Hybrid approaches that combine different spin qubit encodings offer promising advantages. In particular, the additional degrees of freedom available in exchange-only qubits and their extensions enable enhanced spin lifetimes and facilitate error detection through the use of auxiliary spins. We show that integrating Loss–DiVincenzo with exchange-only qubits provides a practical route to realizing these benefits while remaining compatible with spin-shuttling architectures. We further demonstrate how the singlet-only exchange-only qubit can be employed for error detection, and we present a fault-tolerant $\pi$-rotation about each of the three control axes of the (singlet-only) exchange-only qubit. By enabling error detection at the lowest encoding level, our approach effectively converts charge and nuclear noise into erasures, thereby suppressing error propagation and potentially enhancing the performance of quantum error-correction schemes. More broadly, our work establishes a new perspective on the singlet-only exchange-only qubit as a logical encoding, opening the door to fault-tolerant gate constructions and spin-tailored quantum error-correction protocols for semiconductor-based quantum computing.

\end{abstract}

\maketitle

\section{Introduction}
\label{sec:introduction}
Scalable quantum computing requires physical platforms to have large qubit registers with low error rates to preserve quantum coherence. Spin qubits \cite{Burkard_2023} have emerged as a particularly attractive candidate in this context, owing to their long coherence times and compatibility with established semiconductor technologies. A canonical example is the Loss–DiVincenzo qubit~\cite{PhysRevA.57.120}, in which quantum information is encoded in the spin-up and spin-down states of a single electron confined in a gate-defined semiconductor quantum dot, where electrostatic potentials created by external electrodes form tunable traps within a semiconductor device. Initialization and readout are typically achieved through spin-selective tunneling \cite{Elzerman_2004}, while single-qubit and two-qubit operations can be realized through combinations of local static and time-dependent control fields \cite{Koppens_2006, doi:10.1126/science.1116955}. Various spin-qubit platforms have demonstrated gate fidelities exceeding 99\%, and small-scale quantum processors have been realized \cite{Hendrickx_2021, Philips_2022, Mills_2022, Noiri_2022, f285-l2v5, undseth2026weightfourparitycheckssilicon}.

Control protocols based on strong external driving fields can be relatively slow and can cause overheating in devices, which in turn can enhance decoherence \cite{Takeda_2018, Zwerver_2022, doi:10.1126/sciadv.add9408, 1fz7-tjq8}. An appealing alternative for the control is provided by the Heisenberg exchange interaction, which enables much faster gate operations \cite{Medford_2013, Andrews_2019, Weinstein_2023, M_dzik_2025, Acuna_2024, njq3-fcdd}. Moreover, exchange interactions alone are sufficient for universal quantum computation when quantum information is encoded in suitable multi-spin subspaces where the total spin quantum numbers do not change \cite{PhysRevLett.85.1758, PhysRevLett.85.3520, PhysRevA.63.042307}. This idea is realized in the exchange-only (EO) qubit \cite{DiVincenzo_2000}, where an effective qubit is encoded in the collective state of three spins. In this architecture, quantum gate operations are implemented solely by sequentially and selectively pulsing the coupling between neighboring spins on and off, eliminating the need for direct spin manipulation by oscillating magnetic fields. Recent advances in the integration and operation of multiple exchange-only qubits on a single chip highlight their significant potential for quantum computation~\cite{M_dzik_2025, membersofthehrlquantumteam2026digitallycontrolledsiliconquantum}.

\begin{figure}[ht]
    \centering
    \includegraphics[width=0.99\linewidth]{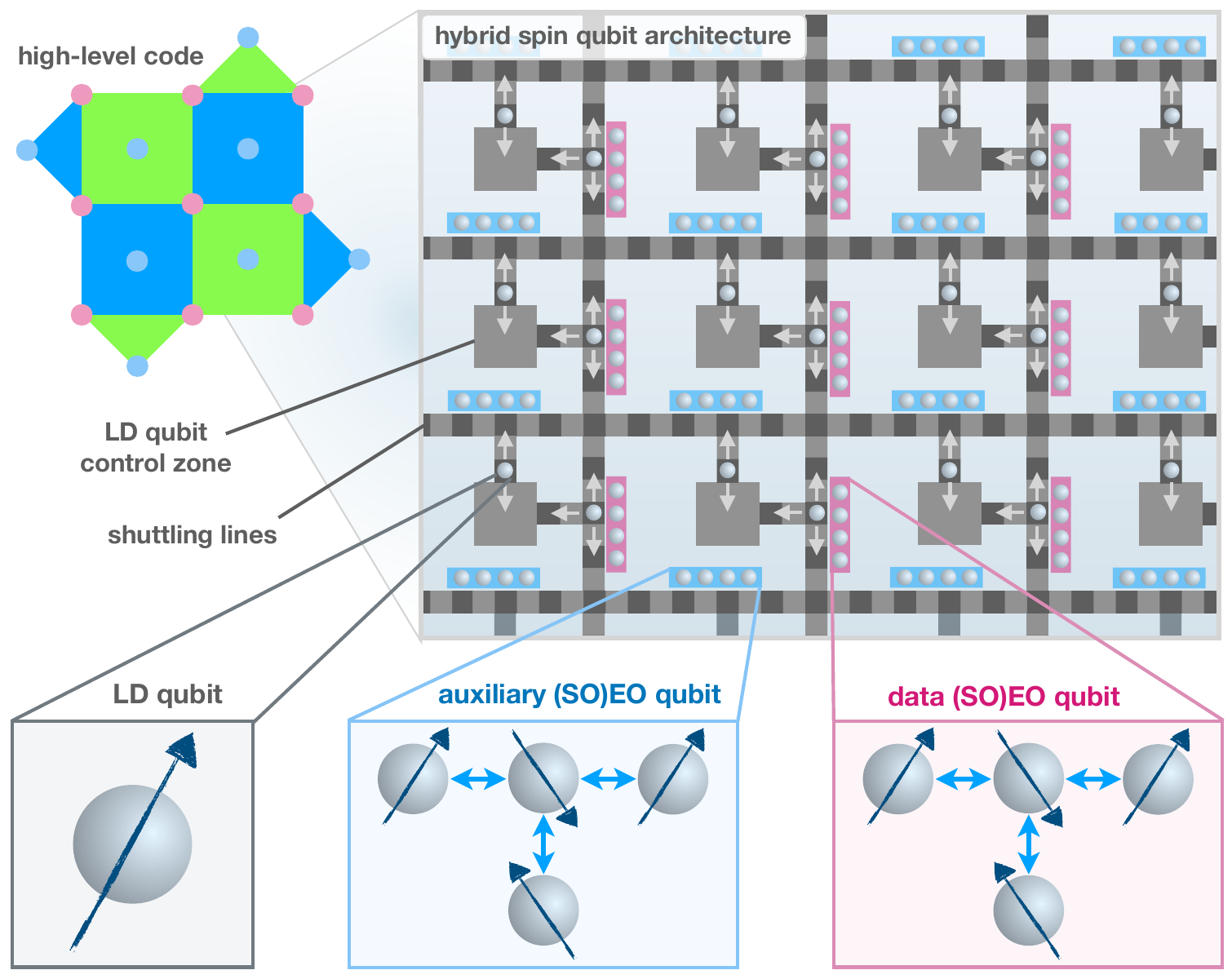}
    \caption{Schematic of a shuttling-based hybrid spin qubit architecture to implement quantum error correction with a high-level code, e.g. the surface code, where data and auxiliary qubits of the surface code are (singlet-only) exchange-only qubits. The blue double-headed arrows indicate exchange interactions between spins. (SO)EO qubits can travel across the chip via shuttling lines. In fact, any spin can be shuttled across the chip, enabling arbitrary connectivity. Additional auxiliary Loss-DiVincenzo (LD) qubits can be used for error detection in the (SO)EO encoding via shuttling between (SO)EO qubits and LD control zones where single-spin rotations are possible.}
    \label{fig:Figure1}
\end{figure}

For the practical use of quantum computation, a large number of reliable qubits is required. High-precision operations, in turn, rely on protection protocols against errors in qubits and gates. After characterizing the hardware and noise sources, error mitigation techniques such as dynamical decoupling can be applied to enhance the qubit performance \cite{Muhonen_2014, Laucht_2016, PhysRevB.88.161303, West_2012, PRXQuantum.5.020356}. In addition to mitigation, to reach low logical error rates required for practical quantum computation, quantum error correction will be essential, relying on the encoding of logical qubits into many physical qubits and the implementation of fault-tolerant operations~\cite{gottesman2009introductionquantumerrorcorrection, aharonov1999faulttolerantquantumcomputationconstant, Knill_1998}. The protection against noise is realized by the readout of stabilizer operators, which correspond to additional degrees of freedom in the encoding and allow for error detection and correction of errors, based on observed measurement outcomes. 
Accordingly, research on quantum error correction with spin qubits is gaining increasing attention~\cite{PhysRevA.109.032433, PhysRevB.110.075302, 4p4q-fm3y, PhysRevA.110.012417, yenilen2025performancespinqubitshuttling, siegel2026erasureconversionsinglettripletspin, chadwick2026manufacturablesurfacecodearchitecture, gutierrez2025comparison, undseth2026weightfourparitycheckssilicon}. 
To date, only limited efforts have been made to integrate EO qubits with quantum error detection or correction schemes~\cite{PhysRevA.63.042307, PhysRevLett.85.1758, membersofthehrlquantumteam2026digitallycontrolledsiliconquantum}. Current approaches rely on existing codes where each data qubit corresponds to an EO qubit \cite{chadwick2026manufacturablesurfacecodearchitecture}. However, the additional degree of freedom in exchange-only qubits and its extensions allows for at least error detection and reset-if-leaked procedures as proposed in Ref.~\cite{langrock2020resetifleakedprocedureencodedspin}. Catching errors at this low-level encoding will prevent errors from spreading across a larger quantum code by resetting the qubit information and re-encoding into the higher-level code.  In this paper, we investigate mitigation techniques as well as fault-tolerant error-detection protocols which can increase the qubit performance using additional single-spin qubits that can be available on spin-qubit architectures using, e.g., spin shuttling techniques \cite{PRXQuantum.4.030303, Mills_2019, Seidler_2022, De_Smet_2025, undseth2026weightfourparitycheckssilicon} as depicted in Fig.~\ref{fig:Figure1}.
Combining different types of spin qubits can be advantageous to tailor the architecture components to different purposes and hybrid gates between them are required to transfer information \cite{gutierrez2025comparison, Mehl_2015}. 

In this paper, we demonstrate the potential of using the additional degrees of freedom in the exchange-only qubit encoding. We investigate a natural extension of the EO qubit to the singlet-only exchange-only (SOEO) qubit by adding a fourth spin to the encoding, and show how a hybrid spin qubit architecture can benefit from error detection on the lowest encoding level.
This paper is organized as follows. 
In Section~\ref{sec:JJJgate}, we investigate a way to address the additional degree of freedom in an EO qubit using an additional spin that interacts only via exchange interaction. 
In Section~\ref{sec:results-SOEO}, we combine the SOEO qubit with two LD qubits into a hybrid spin qubit architecture to enable error detection using fault-tolerant stabilizer readout and investigate the performance under circuit level noise. We discuss the advantage of our protocol for error correction in Section~\ref{sec:discussion}.
In Section~\ref{sec:basics}, we summarize the EO and SOEO qubit and formulate their native operations in a stabilizer framework. We also briefly review the concept of fault tolerance and stabilizer readout. 


\section{Results} \label{sec:results}
\subsection{Addressing the exchange-only gauge qubit} \label{sec:JJJgate}
\begin{figure}[ht]
    \centering
    \includegraphics[width=0.95\linewidth]{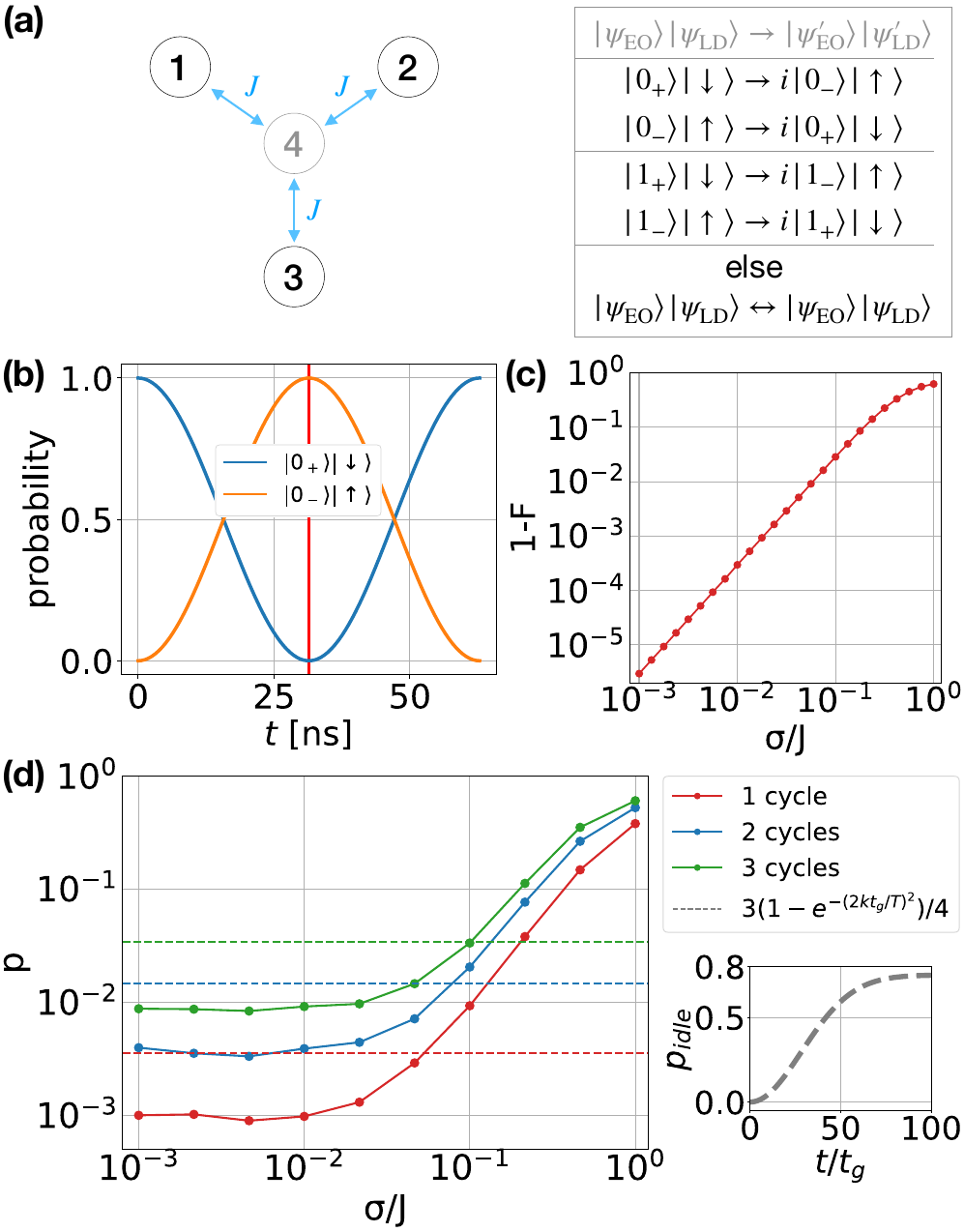}
    \caption{(a)~Simultaneous exchange pulse between each EO qubit spin (1, 2 and 3) and one LD qubit (4) leads to an exchange between the gauge information and the LD spin projection. (b)~Measurement probability of states $\ket{0_{+}}\ket{\downarrow}$ and $\ket{0_{-}}\ket{\uparrow}$ for the initial state $\ket{0_{+}}\ket{\downarrow}$ evolving in time according to Eq.~\eqref{eq:UJJJ}. The red vertical line in the middle corresponds to the gate time $t_g=\pi/J$, where $J=100\,$MHz. (c)~Infidelity depending on relative noise on the exchange interaction strength for the gate $U(t_g)$ in Eq.~\eqref{eq:UJJJ-gate}. (d) Error probability $p$ after $k$ repetitions of performing the unitary $U(t_g)$ with relative gate noise $\sigma/J$ followed by a measurement of the LD qubit in the $z$-basis. The error is averaged over the input states $\ket{\psi_{\rm in}} \in \{\ket{0_-}, \ket{1_-}, \ket{0_-}\pm\ket{1_-}, \ket{0_-} \pm i\ket{1_-}\}$, which suffer from single-qubit depolarizing noise with probability $p_{\rm idle}$ (bottom right) and $T=40t_g$. Horizontal lines correspond to the idling case during the respective cycle time.}
    \label{fig:JJJgate}
\end{figure}

The exchange-only qubit consists of three electrons in three quantum dots and is operated relying solely on the exchange interactions between them. It encodes two qubits, a primary qubit and a gauge qubit that usually remains unused. We summarize the qubit encoding and its native operations in Section~\ref{sec:methods-EO}. It is possible to transfer the gauge information of an exchange only-qubit, i.e. whether it is in the $S_z=+1/2$ or $S_z=-1/2$ subspace, to an LD qubit and vice versa using simultaneous exchange pulses as shown in Fig.~\ref{fig:JJJgate}a.
Here, the LD qubit is the middle spin labeled by 4 and the exchange-only qubit contains spins 1, 2 and 3. When all three shown exchange interactions are simultaneously switched on to the same value $J$, the time evolution is given by
\begin{align}
    U(t) = e^{-iJ (\mathbf{S}_1\cdot\mathbf{S}_4 + \mathbf{S}_2\cdot\mathbf{S}_4 + \mathbf{S}_3\cdot\mathbf{S}_4 -3/4)t}, \label{eq:UJJJ}
\end{align}
with $\mathbf{S}_i=(\sigma_x,\sigma_y,\sigma_z)/2$, in units where $\hbar=1$.
For a gate time $t_g=\pi/J$, this results in an iSWAP transition between the LD spin degree of freedom and the gauge qubit $S_z=\pm 1/2$ of the EO qubit. We write the unitary in terms of the qubit states $\ket{0_\pm}$ and $\ket{1_\pm}$, respectively, 
\begin{align}
    \begin{split}
    U(t_g) = &i \ket{0_-} \ket{\uparrow}\bra{0_+} \bra{\downarrow} 
    + i \ket{1_-} \ket{\uparrow}\bra{1_+} \bra{\downarrow}\\
    + &i \ket{0_+} \ket{\downarrow}\bra{0_-} \bra{\uparrow} 
    + i \ket{1_+} \ket{\downarrow}\bra{1_-} \bra{\uparrow}\\
    + &\ket{0_-} \ket{\downarrow}\bra{0_-} \bra{\downarrow} 
    + \ket{1_-} \ket{\downarrow}\bra{1_-} \bra{\downarrow}\\
    + &\ket{0_+} \ket{\uparrow}\bra{0_+} \bra{\uparrow} 
    + \ket{1_+} \ket{\uparrow}\bra{1_+} \bra{\uparrow}\\
    + &\sum_{i=1}^4\sum_{ s\in\{\uparrow, \downarrow\}} \ket{L_i} \ket{s}\bra{L_i} \bra{s},
    \end{split}\label{eq:UJJJ-gate}
\end{align}
where $\ket{L_i}$ correspond to the leakage states of the EO qubit (see Section~\ref{sec:methods-EO}).
The transitions are also shown next to the schematic connectivity in Fig.~\ref{fig:JJJgate}a. 

In Fig.~\ref{fig:JJJgate}b, we show the measurement probability $\lvert\bra{\psi} U(t) \ket{0_{+}}\ket{\downarrow}\rvert^2$ of states $\ket \psi = \ket{0_{+}}\ket{\downarrow}, \ket{0_{-}}\ket{\uparrow}$ for the initial state $\ket{0_{+}}\ket{\downarrow}$ evolving in time under Eq.~\eqref{eq:UJJJ}. After gate time $t_g$ (vertical red line) the gauge information and the LD qubit state have swapped. To estimate the realistic performance of such a gate, we assume each of the three exchange interactions fluctuates independently. For simplicity, we assume quasi-static noise with standard deviation $\sigma$ and perform a Monte Carlo simulation with $10^4$ runs to compute the infidelity as a function of $\sigma/J$, as shown in Fig.~\ref{fig:JJJgate}c.

A useful protocol for exchange-only qubits encoded in the $\{\ket{0_-}, \ket{1_-}\}$ subspace rather than the subsystem could be as follows:
(i)~Initialize the LD qubit in $\ket{\downarrow}$,
(ii)~apply the evolution $U(t_g)$, (iii)~measure the LD qubit. In App.~\ref{App:JJJ-map}, we show that this results in
\begin{align}
    \rho \rightarrow A_0 \rho A_0^{\dagger}
    + A_1 \rho A_1^{\dagger},
\end{align}
where $A_{0} = \ketbra{0_-}{0_-} + \ketbra{1_-}{1_-} + \sum_{i=1}^{4} \ketbra{L_i}{L_i} $ and $A_{1} = i \ketbra{0_-}{0_+} + i \ketbra{1_-}{1_+}$. 
That is, if at the end of the protocol one measures the LD qubit in $\ket{\downarrow}$, we have not acted on the EO qubit, as it was in the space spanned by $\{\ket{0_-}, \ket{1_-}\}$ or it was leaked. Otherwise, if we measure the LD qubit in $\ket{\uparrow}$, the EO qubit was at least partially in the subspace spanned by $\{\ket{0_+}, \ket{1_+}\}$ and is now projected onto the $\{\ket{0_-}, \ket{1_-}\}$ subspace. 

We simulate such an experiment in Fig.~\ref{fig:JJJgate}d. For this, we use an input state $\ket{\psi_{\rm in}} \in \{\ket{0_-}, \ket{1_-}, \ket{0_-}\pm\ket{1_-}, \ket{0_-} \pm i\ket{1_-}\}$, which suffers from a noise process $\mathcal E$ while idling. We model this noise by applying a single-spin depolarizing Pauli error occurring with probability $p_{\rm idle}(t_g)$ on any single spin with $p_{\rm idle}(t_g)= 3(1-e^{-(t_g/T)^2})/4$, with $T =40t_g$ (bottom right plot). This effective error model is derived in Section~\ref{sec:error-model}. The error is followed by a noisy $U(t_g)$ gate and perfect measurement of the auxiliary LD qubit. The gate noise is modeled as quasi-static deviation of the exchange value $J$ with standard deviation~$\sigma$. We perform multiple iterations of the idling followed by $U(t_g)$, and calculate the EO qubit error probability $p = 1-\lvert\bra{\psi_{\rm in}} (P_{m_k} U(t_g) E_k) \dots (P_{m_1} U(t_g) E_1) \ket{\psi_{\rm in}}\rvert^2$, where $P_{m_i}$ is the projector of the LD qubit with respective measurement outcome $m_i=\downarrow, \uparrow$ and $E_i$ is the composition of the idling noise in the $i$th cycle. In Fig.~\ref{fig:JJJgate}d, we show the EO qubit error probability $p$ depending on the signal-to-noise ratio $\sigma/J$ for $k=1,2,3$ cycles averaged over all runs and over the different input states. For comparison, we show the EO error probability for the same input states, that suffer from single-spin depolarizing noise with probability  $p_{\rm idle} = 3(1-e^{-(2kt_g/T)^2})/4$, without any gates (straight dashed lines). We find that applying $U(t_g)$ enhances the EO qubit lifetime if the gate noise is low enough. This is consistent with the quantum Zeno effect, which slows down the state's evolution by frequent projection onto the correct subspace (see Section~\ref{sec:zeno} for more details). 

Another potential application of the gate $U(t_g)$ is to store additional information that is transferred from an LD qubit to the EO gauge qubit. We do not pursue this possibility further in the present work.

Although we have shown enhancement of the EO qubit lifetime by adding only a single LD qubit, while still relying only on the exchange interaction, realizing the protocol on a real device comes with challenges. First, the LD qubit must be initialized by a nearby reservoir and brought to a region where it can interact with all three exchange couplings simultaneously with the same strength. Although it is in principle possible to achieve such an exchange interaction \cite{Broz2026}, it requires high-precision calibration to realize the star-shaped coupling in a real device. 

\subsection{SOEO performance enhanced by error detection} \label{sec:results-SOEO}
\begin{figure*}[ht]
    \centering
    \includegraphics[width=1\linewidth]{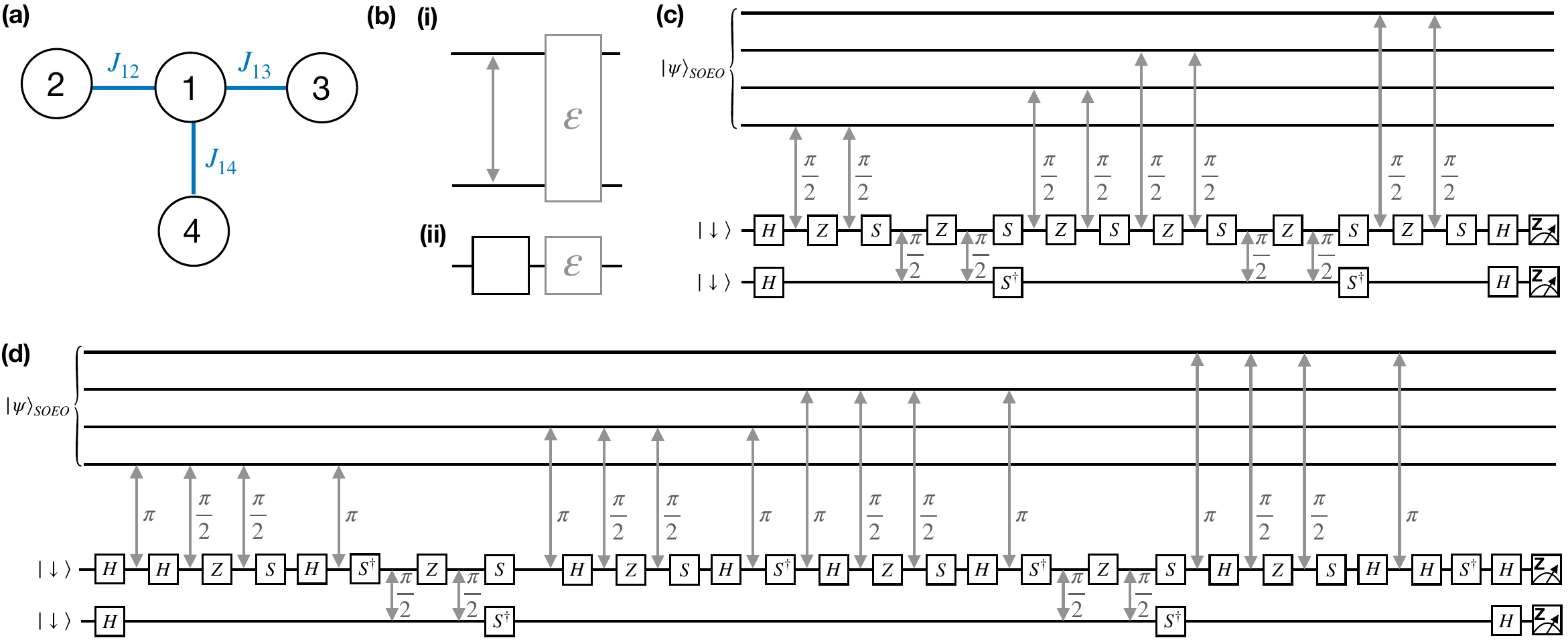}
    \caption{
    (a) SOEO qubit with exchange couplings $J_{ij}$. (b) Error model for circuit components. (c) Fault-tolerant $ZZZZ$ and (d) $XXXX$ measurements using the relations in Fig.~\ref{fig:CNOTbetweenSpins}.}
    \label{fig:SOEO-FT-QED-cicuit}
\end{figure*}
Stabilizer readout is an essential ingredient of quantum error detection and correction, as it enables the identification of errors through syndrome measurements. Since the exchange-only (EO) qubit can be confined to a fixed total-spin subspace, measurements of the three-spin $XXX$ or $ZZZ$ operators are, in principle, possible (tensor products are omitted for brevity). However, because these operators do not commute, they cannot be measured simultaneously. In contrast, stabilizers containing an even overlap of $X$ and $Z$ operators do commute and can therefore be measured concurrently. This observation naturally motivates the extension of the EO qubit by incorporating an additional spin into the encoding~\cite{PhysRevLett.85.1758, PhysRevB.95.241303, j1hn-k5m8}. One such extension is the singlet-only exchange-only (SOEO) qubit, which encodes the logical qubit in the $S=0$ subspace (with $S$ the total spin quantum number).
Thus, as a next step, we investigate the SOEO qubit and show how error detection can increase the qubit performance and lifetime. Error detection is a nondeterministic protocol that relies on postselection of runs with a trivial syndrome outcome.

We consider the SOEO qubit as a logical qubit encoding. The stabilizers of the SOEO qubit states, which contain the four-spin operators $XXXX$ and $ZZZZ$, are discussed in more detail in Section~\ref{sec:methods-SOEO}. In order to read out these two stabilizer elements, CNOT gates between single spins are required. Note that in a SOEO qubit, single-spin rotations are unavailable. However, for the stabilizer readout circuit, we will use two auxiliary spins, one for recording the measurement outcome, and one (the flag qubit) for making the readout fault tolerant \cite{Chao_2018, Chamberland_2018, Chao_2020}. These auxiliary qubits we will realize as LD qubits, for which single-qubit gates are available. We assume that these LD qubits can be shuttled between the SOEO qubit and an interaction zone, where single-qubit rotations can be performed (see Fig.~\ref{fig:Figure1} and App.~\ref{App:shuttling}).
In Section~\ref{sec:methods-SOEO}, we discuss the construction of the fault-tolerant readout circuits that are shown in Figs.~\ref{fig:SOEO-FT-QED-cicuit}c and d. The double arrows depict $\sqrt{\rm SWAP}$ and SWAP operations, which correspond to $\pi/2$ and $\pi$ pulses of the exchange interaction between the two spins.

\begin{figure*}[ht]
    \centering
    \includegraphics[width=0.80
    \linewidth]{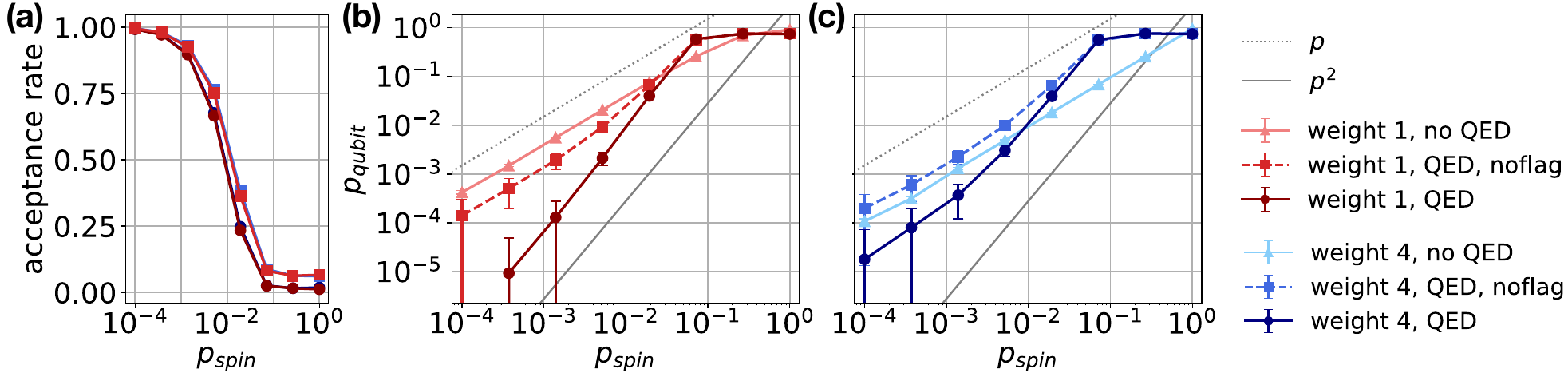}
    \caption{State simulation of the fault-tolerant error detection circuit in Fig.~\ref{fig:SOEO-FT-QED-cicuit}c and d averaged over the  $X$, $Y$, and $Z$ eigenstates of the encoded qubit. Noise is modeled as depolarizing noise on the gates plus an incoming weight-one (red tones) or weight-four (blue tones) fault followed by a noiseless round of error detection. The circuit was simulated with and without the lower flag qubit and the respective gates. (a) Acceptance rates with and without flag. (b) Success probabilities depending on the spin error $p_{\rm spin}$ for incoming weight-one faults which correspond to the acceptance after the noiseless detection step. The error probability is shown without QED in light red, with QED without the flag in red, and with QED and with flag in dark red. The gray dashed and solid lines show a linear and quadratic scaling. (c) Error probability without QED in light blue, with QED but without the flag in blue, and with QED and with flag in dark blue.}
    \label{fig:SOEO-QED-simulation}
\end{figure*}

We execute the stabilizer measurement using circuit-level noise as follows. We first assume a perfect initialization of both the SOEO qubit $\ket{\psi}_{\rm SOEO} = \ket{\psi_{\rm in}}$ and the two auxiliary spins $\ket{\downarrow \downarrow}$, which is followed by a depolarizing noise channel [Eq.~\eqref{eq:error-on-single-qubit-gates}] with probability $p_{\rm spin}$ on each of the six spins individually. We simulate the circuits in Figs.~\ref{fig:SOEO-FT-QED-cicuit}c and d, assuming single-spin depolarizing noise on single-spin gates and two-spin depolarizing noise on two-spin operations [Eq.~\eqref{eq:error-on-two-qubit-gates}] (both likewise with $p_{\rm spin}$). Measurement errors are taken into account by a noisy Hadamard gate followed by a perfect measurement in the $Z$ basis, which is equivalent to a noisy measurement of the LD qubits in the $X$ basis. To verify fault tolerance, we follow the noisy QED cycle with a noiseless error-detection cycle and confirm that a single physical fault during the protocol cannot produce an undetected logical error \cite{aliferis2005quantumaccuracythresholdconcatenated, preskill1997faulttolerantquantumcomputation}. We run simulations for the qubit states $\ket{\psi_{\rm in}} \in \{\ket{0}, \ket{1}, \ket{+}, \ket{-}, \ket{i}, \ket{-i}\}$ and postselect on all runs with trivial measurement outcomes in both the noisy and noiseless detection rounds. 

The acceptance rate is shown in Fig.~\ref{fig:SOEO-QED-simulation}a for the circuit with (dark red) and without (red) the lower flag spin in Fig.~\ref{fig:SOEO-FT-QED-cicuit}c and d. For the lowest error rate, a Monte Carlo simulation of up to $10^5$ runs is performed. From the postselected runs, we calculate the qubit error probability as $p_{\rm qubit} = 1 - \lvert\braket{\psi_{\rm in}}{\psi}\rvert^2$ which, after averaging over all runs, is shown in Fig.~\ref{fig:SOEO-QED-simulation}b. For comparison, we calculate the error probability for the SOEO qubit state suffering only from an incoming error, without using QED. We find that the error probability follows a linear scaling in $p_{\rm spin}$ when no QED is applied. Using QED without the flag seems to slightly decrease the error rate $p_{\mathrm{qubit}}$ in an interval that lies below $p_{\rm spin}=0.02$ and also follows a linear scaling. On the other hand, the error probability due to the flagged circuit clearly follows a quadratic behavior manifesting that the circuit is fault tolerant, and we obtain a pseudo-threshold of $p_{\rm spin}=3.5\%$. We thus conclude that for spin error rates below the threshold value, the measurement circuit does not introduce more errors than it can detect.

So far, we have assumed single-spin depolarizing noise as the incoming noise model, which would be a reasonable assumption, provided that fault tolerance is maintained throughout all preceding operations in the quantum computation prior to the QED cycle. However, when operating the SOEO qubit using its native exchange pulses, errors on the spins of the SOEO qubit can occur with arbitrary weight. As a worst-case scenario, we therefore assume a four-spin depolarizing noise channel as an incoming error channel, followed by the stabilizer readout circuits in Figs.~\ref{fig:SOEO-FT-QED-cicuit}c and d and a subsequent round of noiseless QED. The acceptance rate in Fig.~\ref{fig:SOEO-QED-simulation}a for the circuit with (dark blue) and without (blue) the flag spin does not deviate much from the ones for weight-one faults. 

In Fig.~\ref{fig:SOEO-QED-simulation}c, we show the error probability $p_{\mathrm{qubit}}$ resulting from the incoming four-qubit depolarizing noise without the QED round (light blue), with QED but without the flag (blue), and with flagged QED (dark blue). The bare qubit error probability without QED follows a clear linear trend. When comparing this to the error rates in the case with incoming weight-one faults, we observe a slightly different behavior. While generally any weight-one fault leads to a qubit error, some weight-four errors correspond to a stabilizer. Thus, not each incoming error ultimately corresponds to a qubit error. 
We further find that the unflagged QED performs worse compared to not measuring the stabilizers and approaches the linear scaling for low error rates. Finally, our calculations indicate that flagged QED can substantially reduce the qubit error by up to one order of magnitude below a spin error rate of $p_{\rm spin} = 0.9$\%. Since the stabilizer readout circuit is not able to detect all incoming errors, we do not obtain a quadratic scaling. However, for spin error rates $p_{\rm spin} \in [10^{-3}, 10^{-1}]$ the qubit error rate follows a quadratic behavior, and approaches a linear scaling for $p_{\rm spin} \ll 10^{-3}$. We note that although some errors remain undetected, the fault-tolerant stabilizer readout reduces the qubit error rate for any error weight. Although over-rotations in single-qubit gates are generally not detected, over-rotations during two-qubit gate sequences that lead to leakage are expected to be detectable. 

\subsubsection{Memory experiment}\label{sec:memory}

\begin{figure}[ht]
    \centering
    \includegraphics[width=0.99\linewidth]{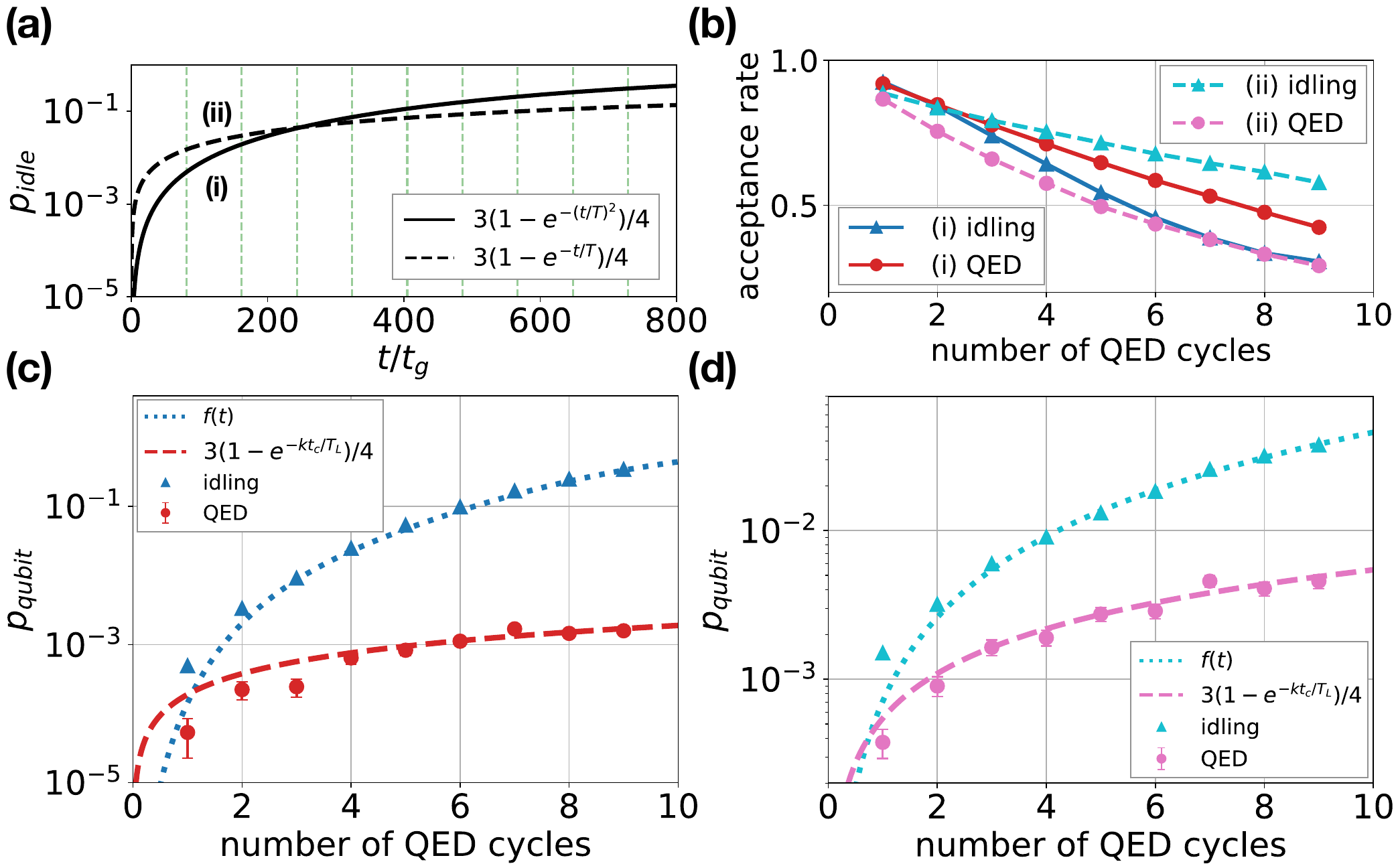} 
    \caption{Memory experiment simulation over a varying number of quantum error detection cycles. (a) The two types of idling probability decay (i)~$p_{\rm idle}(t) = 3(1-e^{-(t/T)^2})/4$, and (ii)~$p_{\rm idle}(t) = 3(1-e^{-t/T})/4$. (b) Acceptance rate for type-(i) decay without QED (blue) and with QED (red), and similarly for type-(ii) decay in cyan and pink.  (c) Qubit error rate over a varying number of QED cycles for type-(i) decay. The simulations including the QED cycles are shown in red and are fitted as described in the main text. Note that although the error model is type-(i), the fit is to an effective type-(ii) error model in the number of cycles. The comparable idling experiment, without QED cycles, is shown in blue, where the fitting function is detailed in the main text. (d) Similar qubit error rate for type-(ii) decay, with different fits.} 
    \label{fig:SOEO-QED-over-many-cycles}
\end{figure}

As a next step, we simulate a quantum memory experiment, in which we include idling errors. We assume that all single-spin gates and all $\pi/2$ exchange pulses have gate time $t_g$ and have a depolarizing error probability $p_{\rm spin} =0.001$. We will use the time interval $t_g$ for time units in the remainder of this section. To model idling, we employ the two noise models derived in Section~\ref{sec:idling_noise}, which capture qualitatively different error behaviors that depend on the noise spectrum of the environment. Per spin, these models are a depolarizing channel with error probability (i)~$p_{\rm idle}(t) = 3(1-e^{-(t/T)^2})/4$ with $T = 1 \cdot 10^{3}t_g$, or (ii)~$p_{\rm idle}(t) = 3(1-e^{-t/T})/4$ where $T = 4\cdot 10^{3} t_g$. The characteristic times $T$ are chosen such that both models exhibit a comparable decay over the time window of interest (see Figs.~\ref{fig:SOEO-QED-over-many-cycles}a, c and d); their precise values are not essential for the conclusions drawn below. They are not intended to model a particular device, but serve as representative coherence scales broadly motivated by experiments on silicon spin qubits~\cite{Veldhorst_2014, Steinacker2025, Weinstein_2023}. 
We start the simulation with an initial single-spin depolarizing channel on all four spins, with error probability $p_{\rm idle}(t_g)$, and set the time $t=0$. Next, we repeatedly run the circuits in Fig.~\ref{fig:SOEO-FT-QED-cicuit}. For the $ZZZZ$ and $XXXX$ readout in Figs.~\ref{fig:SOEO-FT-QED-cicuit}c and d, all spins suffer from an idling error with probability $p_{\rm idle}(t)$ while they are not touched by a gate. During a gate, we assume a gate error $p_{\rm spin}$. After a gate acting on a spin, the spin continues suffering from idling errors until the next gate or the end of the circuit as described in Section~\ref{sec:idling_noise}. The whole simulation is followed by perfect $XXXX$ and $ZZZZ$ stabilizer readout. The results are averaged over input states $\ket{\psi_{\rm in}} \in \{\ket{0}, \ket{1}, \ket{+}, \ket{-}, \ket{i}, \ket{-i}\}$. 

The acceptance rates are shown in Fig.~\ref{fig:SOEO-QED-over-many-cycles}b. The qubit error probabilities are shown in Figs.~\ref{fig:SOEO-QED-over-many-cycles}c and d as red and pink data points. They are both fitted to a decay of the form $(1-e^{-rt_c/T_L})/2$, with (c)~$T_L = 3.20 \cdot 10^{5} t_g$ and (d)~$T_L = 1.11 \cdot 10^{5} t_g$. Here, $T_{L}$ denotes the logical decoherence time, $t_c = 81 t_g$ is the time for one QED cycle, and $r$ is the number of cycles.

We compare the QED protocol to a pure idling situation. As before, we start with an initial single-spin  depolarizing channel on all four spins. Then, we introduce an idling error $p_{\rm idle}(t)$, with $t$ the total duration of the respective circuit with a varying number of QED cycles
and perform a noiseless QED cycle at the end. In Fig.~\ref{fig:SOEO-QED-over-many-cycles}b, the acceptance rates are shown in blue and cyan. The logical error probabilities are shown as blue and cyan triangles in Figs.~\ref{fig:SOEO-QED-over-many-cycles}c and d. In Figs.~\ref{fig:SOEO-QED-over-many-cycles}c and d, we fit the idling decay to a function $f(t) = \alpha p^2(1-p)^2 + \beta p^3(1-p) + \gamma p^4$ with $p = 3(1-e^{-(rt_c/\tau)^2})/4$ for (c) and $p = 3(1-e^{-rt_c/\tau})/4$ for (d). We discuss the error probability and the fitting ansatz in App.~\ref{App:FitErrorProbability}. We find (c) $\alpha = 0.83$, $\beta = 0.06$, $\gamma = 2.59$, $\tau \approx 6.06\cdot 10^2 t_g$, and (d) $\alpha = 0.33$, $\beta = 0.02$, $\gamma = 1.09$, $\tau \approx 1.25\cdot 10^3 t_g$. We note that, for the data points corresponding to one QED cycle, initialization errors are more dominant than for later data points, which explains the deviation from the fitted line. In both cases we observe a qubit lifetime enhancement when applying the QED cycle.

In Fig.~\ref{fig:SOEO-QED-over-many-cycles}b the acceptance rate of the two types of spin decays, both with QED and idling simulations, are shown. The solid lines correspond to the type-(i) decay from Fig.~\ref{fig:SOEO-QED-over-many-cycles}a. For this type of idling noise, we find that the QED acceptance rate (red solid) is larger than for the idling case (blue solid). This indicates that the qubit is possibly stabilized due to the Zeno effect (Section~\ref{sec:zeno}). The results in Fig.~\ref{fig:SOEO-QED-over-many-cycles}c indeed show a substantial reduction in the qubit error rate of up to two orders of magnitude. 
Overall, fewer runs are discarded when using the QED protocol than in the idling case, and the remaining runs exhibit a lower error rate. Thus, the protocol simultaneously improves the retention rate and the quality of the accepted runs.

In Fig.~\ref{fig:SOEO-QED-over-many-cycles}b the dashed lines correspond to the decay from  Fig.~\ref{fig:SOEO-QED-over-many-cycles}a (ii). Comparing the idling simulations (cyan dashed) and the QED simulations (pink dashed), we observe a reduced acceptance rate when using QED. At the same time, Fig.~\ref{fig:SOEO-QED-over-many-cycles}d indicates that the qubit error probability decreases by almost one order of magnitude when using QED. There are no indications of a Zeno effect here, which is the expected behavior for this type of decay that is linear in time around the origin (Section~\ref{sec:zeno}). When using QED, more runs are discarded than in the idling case, and the remaining runs prevail with a lower error rate. The protocol therefore offers improved fidelity of the accepted runs, at the cost of a lower retention rate.

\subsection{Fault-tolerant \texorpdfstring{$\pi$}{pi}-rotations} \label{sec:FT-SWAP}
For our error-detection scheme, we use the stabilizers $XXXX$ and $ZZZZ$, which define the $[[4,2,2]]$ code, encoding two logical qubits into four physical qubits with code distance two (see App.~\ref{App:4-2-2-code}). In the $[[4,2,2]]$ code, single-qubit gates can be applied \textit{transversally}, that is, using physical gates acting on single spins separately. Consequently, weight-two faults occur with probability $p^2$, making transversal gates fault tolerant by construction. However, for the SOEO qubit, 
both single- and two-qubit operations rely on two-spin interactions, which may lead to weight-two faults. As a result, the architecture lacks native transversal single-qubit gates. Therefore, a fault-tolerant gate set has not yet been established for the SOEO qubit.

We will refer to one SOEO qubit as one code block. As soon as one operation touches more than one spin within one code block at a time, it is not considered transversal and is thus not fault tolerant.   

\begin{figure}[ht]
    \centering\includegraphics[width=0.99\linewidth]{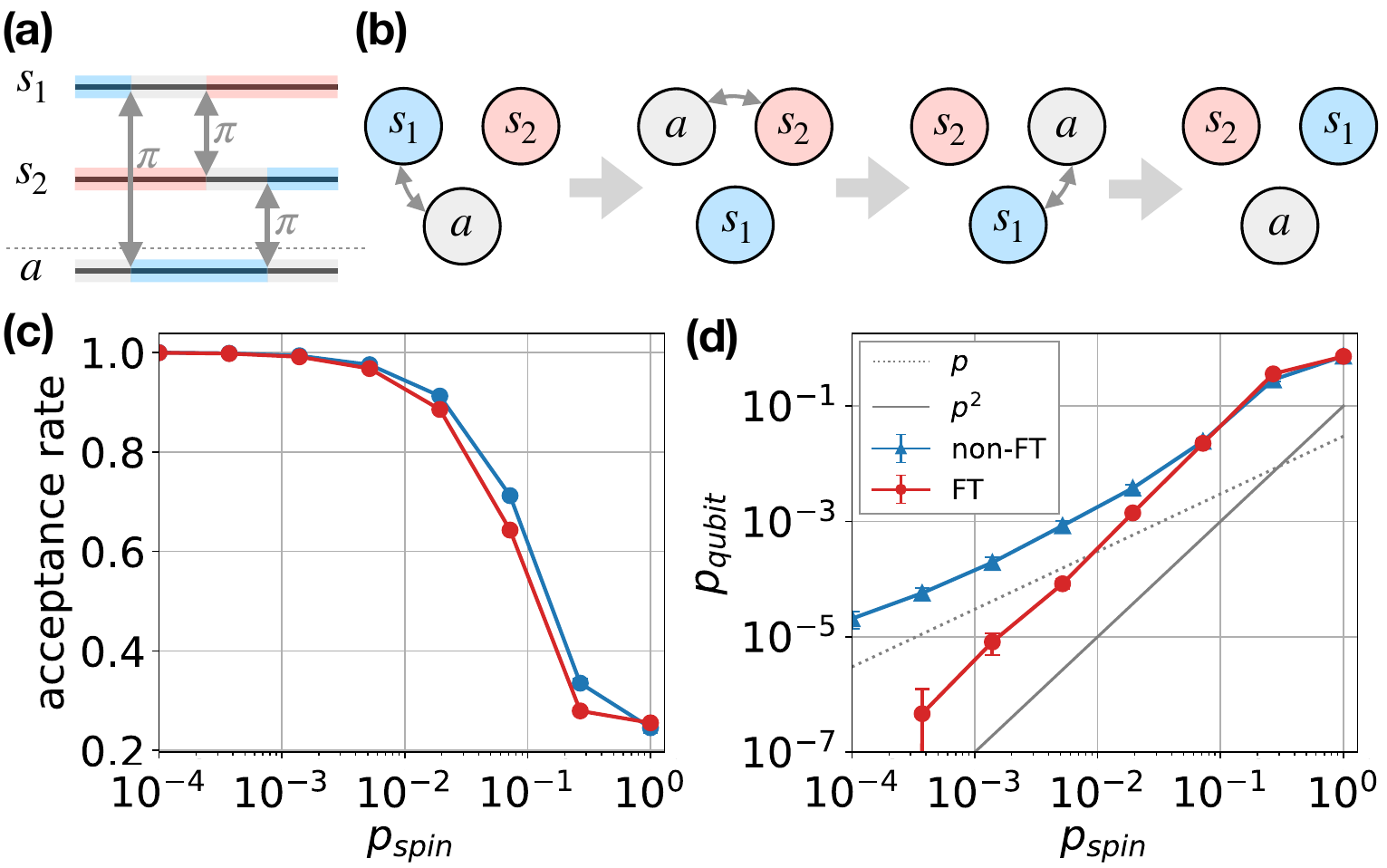}
    \caption{(a) Fault-tolerant SWAP gate between spins $s_1$ and $s_2$ using one auxiliary qubit $a$ in an arbitrary state. (b)~Triangular arrangement of the three spins involved in the fault-tolerant SWAP protocol. (c) Acceptance rate and (d) error probability for the direct $Z_L$ and the fault-tolerant (FT) $Z_L$ gate, corresponding to a SWAP between spins $s_1$ and $s_2$.}
    \label{fig:FTSWAP-all}
\end{figure}

A practical approach to construct fault-tolerant protocols is the use of Clifford gates, as they map Pauli spin errors onto other Pauli spin errors. This makes the error propagation easy to track, which then also allows one to catch any undetectable error. In general, the Clifford group is generated by $\{ H, S, \mathrm{CNOT} \}$.
Exchange pulse sequences used for single-qubit gates are often non-Clifford in the standard spin basis. However, there is one exception, which is a $\pi$-pulse, corresponding to the SWAP operation between two spins. The SWAP gate simply swaps incoming Pauli errors, and is thus a Clifford operation. A SWAP operation can be made fault tolerant by introducing an additional spin \cite{FTSWAP} as shown in Fig.~\ref{fig:FTSWAP-all}a. Figure~\ref{fig:FTSWAP-all}b shows a triangular arrangement in which spins $s_1$ and $s_2$ are exchanged via an auxiliary spin that can be in an arbitrary state. It is important that the two data spins $s_1$ and $s_2$ are never touched by the same SWAP operation.

Consequently, spin SWAP operations can be performed fault tolerantly, provided that an auxiliary spin is available adjacent to the two spins involved in the SWAP.
In the encoded SOEO qubit basis, a $\pi$-pulse on the exchange interactions $J_{12}$, $J_{13}$, and $J_{14}$ in Fig.~\ref{fig:SOEO-FT-QED-cicuit}a corresponds to the logical operations $Z_L$, $N_L$, and $M_L$, respectively, where we assign $N_L = -(\sqrt{3} X_L + Z_L)/2$, and $M_L = (\sqrt{3} X_L - Z_L)/2$ to the $\pi$ rotations around the corresponding vectors $\vec{n}=-(\sqrt{3},0,1)$ and $\vec{m}=(\sqrt{3},0,-1)$. Note that we denote logical $X$ and $Z$ operators of the SOEO qubit as $X_L$ and $Z_L$. For the EO qubit encoded in three spins, the rotation axes are equivalent (see Section~\ref{sec:methods-EO} for more details).
As a result, together with the fault-tolerant state preparation of the logical $\ket{0}$ state discussed in the methods section, the eigenstates of $N_L$ and $M_L$ can also be prepared fault tolerantly.

To demonstrate the advantage of the fault-tolerant spin SWAP, we simulate the protocol for the $Z_L$ gate of the SOEO qubit. The simulation results for the $N_L$ and $M_L$ gates are equivalent. We consider an incoming single-spin depolarizing channel, together with two-spin gate errors as discussed in Section~\ref{sec:error-model}, each with probability $p_{\rm spin}$, and compare the performance to a direct SWAP operation under the same error model. In both cases, the protocol is followed by a noiseless round of error detection. The acceptance rate is shown in Fig.~\ref{fig:FTSWAP-all}c. After postselection, the  overlap of the ideal final state and the actual final state is calculated. The resulting error rate on the SOEO qubit is shown in Fig.~\ref{fig:FTSWAP-all}d. Despite the comparable acceptance rates of the fault-tolerant and non-fault-tolerant $Z_L$, we find a substantial improvement in the error rate when using the fault-tolerant protocol. Instead of the linear scaling in $p_{\rm spin}$, the fault-tolerant protocol scales quadratically. For a depolarizing channel in the range of $p_{\rm spin} \approx 0.01\%-1\%$, we find an error suppression of up to two orders of magnitude.

We note that $Z_L$ is a logical Pauli operator and thus belongs to the logical Clifford group, whereas $N_L$ and $M_L$ are not in the Clifford group. In fact, $N_L$ and $M_L$ are not contained in any finite level of the Clifford hierarchy~\footnote{The $k$th Clifford hierarchy is recursively defined by $\mathcal{C} (k) = \{U\mid\forall P\in \mathcal{P}, UPU^{\dagger} \in \mathcal{C} (k-1) \}$ where $\mathcal{C} (1) = \mathcal{P}$ is the Pauli group and $\mathcal{C} (2)$ is the Clifford group.} \cite{kliuchnikov2013fastefficientexactsynthesis, Forest_2015}. Many well-known stabilizer codes used for practical error correction, such as the Steane code \cite{PhysRevLett.77.793, PhysRevX.11.041058, Postler_2022} and the surface code \cite{bravyi1998quantumcodeslatticeboundary, PhysRevA.86.032324, Google2023, Krinner_2022}, have only transversal and hence fault-tolerant Clifford gates. They either require magic-state injection or must switch to a code that allows for a fault-tolerant implementation of a non-Clifford gate, such as the 3D color code \cite{PRXQuantum.5.020345, Bravyi_2005}, to enable universal fault-tolerant quantum computation. Thus, the SOEO qubit encoding provides a fault-tolerant non-Clifford gate at the cost of using an additional auxiliary spin that carries no logical information. It is not yet clear if the presence of a fault-tolerant non-Clifford gate of this kind can be used for code switching or magic-state injection protocols; nevertheless, we believe this may be an interesting direction to be explored and places the SOEO qubit encoding in a unique position. 

Furthermore, the current approach allows for fault-tolerant SWAPs not only between physical spins, but also between two EO or SOEO qubits, as they only require $\pi$-pulses of the exchange interactions. In App.~\ref{App:shuttling}, we also discuss how fault-tolerant SWAP operations can be taken into account in the architecture design, and how this can be beneficial for logical CNOT gates enabling dynamic SOEO qubit connectivity. 

In fact, instead of physically executing a SWAP between two spins, it can be implemented perfectly by simply relabelling the spins. This idea was explored for CSS codes in Ref.~\cite{koh2026entanglinglogicalqubitsphysical} leading to so-called phantom codes, where relabelling redistributes
existing correlations among subsystems to realize logical operations. However, in order to take advantage of the relabeling, it requires high connectivity between spins, which could be achieved using the layouts discussed in App.~\ref{App:shuttling}.

So far, we could only identify three single-qubit gates and the SWAP gate to be fault tolerant in the SOEO encoding. This set of operators is far from providing a fault-tolerant universal gate set. 
It remains an open question whether it is possible to construct a fault-tolerant CNOT gate between two EO or SOEO qubits, and whether additional single-qubit gates can be made fault tolerant using auxiliary spins. 

\section{Discussion} \label{sec:discussion}
We have presented a protocol for transferring gauge information between an exchange-only qubit and a Loss-DiVincenzo qubit using simultaneous exchange pulses, and evaluated its performance in the presence of exchange fluctuations. Our results indicate that repeated projections onto one subspace can enhance the EO qubit lifetime provided that the gate noise remains sufficiently low, which is in agreement with the quantum Zeno effect. The protocol may also enable the transfer of additional information between the two spin qubit encodings. However, its practical implementation requires extremely precise gate calibration. Although the protocol may be useful in certain settings, the natural extension of the EO qubit to the singlet-only exchange-only qubit further enhances error resilience by enabling the detection of single-spin errors, potentially making it a more practical architecture for future quantum devices.

The stabilizer readout for the SOEO encoding discussed in this work can be concatenated with higher-distance codes. The low-level error detection provides two benefits in such a setting. First, once an error is detected, the location is known before running the high-level stabilizer readout. Thus, instead of dealing with a qubit error, we convert many errors into erasures which are known to increase the threshold of the surface code substantially as they are easier to decode \cite{PhysRevResearch.2.033042, Wu_2022, Colmenarez_2025, d1v7-nctj, chadwick2025erasureminesweeperexploringhybriderasure}. It also reduces the number of higher-level stabilizer checks, which require the application of several CNOT gates between SOEO qubits. Furthermore, the number of harmful leakage errors of both data and auxiliary SOEO qubits is reduced~\cite{brown2020criticalfaultsleakageerrors} by our proposed protocol.
Second, the low-level stabilizer readout projects the system onto one of the following. (i) If the measurement outcome is nontrivial, a weight-one spin error or a detectable multi-spin fault has occurred and is detected. The qubit can then be reset and re-encoded into the higher-level code before errors spread further to other qubits. (ii) If the measurement outcome is trivial, either no error has happened, or one of the 63 dangerous faults among the in total 243 possible multi-weight faults has occurred (see Appendix~\ref{App:FitErrorProbability} for details on remaining errors). On the one hand, the SOEO stabilizer readout either repeatedly projects the state onto the computational subspace and exploits the Zeno effect where applicable. On the other hand, it modifies the noise model such that the remaining errors are generated by weight-two operators, $\langle X_{i}X_{j}, Z_{k}Z_{l} \rangle$, where $i\neq j$, and $k\neq l$, which could be e.g. single-qubit over-rotations. 

A possible layout for concatenating the SOEO encoding with the surface code based on spin shuttling is shown in Fig.~\ref{fig:Figure1}. Stabilizer readout of the SOEO qubits can be realized using additional spins that can be shuttled between the SOEO qubit and LD control zones. The surface code stabilizer readout can be performed using shuttling of auxiliary SOEO qubits. 

Finally, the fault-tolerant SWAP operation between spin pairs enables not only a fault-tolerant implementation of the $Z_L$ gate, but also of two non-Clifford gates, $N_L$ and $M_L$, which are not fault tolerant in conventional quantum error-correcting codes. This observation opens a potential pathway towards constructing universal fault-tolerant gate sets based on novel code implementations. In particular, it may be explored in the context of magic-state injection and integrated into hybrid spin-qubit architectures. Furthermore, algorithms specifically tailored to such extended gate sets could benefit from the direct availability of these fault-tolerant non-Clifford operations.

\section{Methods} \label{sec:basics}
\subsection{The exchange-only qubit} \label{sec:methods-EO}
The universal control of a qubit using only the exchange interaction requires at least three spins. The EO qubit \cite{DiVincenzo_2000} consists of three electrons each sitting in adjacent quantum dots, where the exchange interaction between two neighboring electrons can be controlled via voltages on the according gate electrodes. These pairwise interactions follow the Heisenberg Hamiltonian
\begin{align}
    H = \sum_{i , j} J_{ij} (t) \left( \mathbf{S}_i \cdot \mathbf{S}_j -\frac{1}{4} \right) , \label{Eq:Hamiltonian}
\end{align}
with $J_{ij}$ the exchange interaction strength between spins $i$ and $j$. A linear arrangement enables universal control via two rotation axes on the Bloch sphere, the $z$-axis and the vector $\vec{n}=-(\sqrt{3},0,1)$ \cite{DiVincenzo_2000, Andrews_2019}, while a triangular arrangement enables an additional rotation axis $\vec{m}=(\sqrt{3},0,-1)$~\cite{Acuna_2024, njq3-fcdd} which is not required for universal control.

We transform the single-qubit Hamiltonian into a basis using the total spin $S$, its projection $S_z$, and the combined spin on sites 1 and 2, $S_{12}$. We then express the Hamiltonian in terms of the basis states $|S S_z S_{12}\rangle$,
\begin{equation}
    \begin{aligned}
    |0_-\rangle &= \left| \frac{1}{2}, -\frac{1}{2}, 0 \right>, & |1_-\rangle &= \left| \frac{1}{2}, -\frac{1}{2}, 1 \right>,\\
    |0_+\rangle &= \left| \frac{1}{2}, +\frac{1}{2}, 0 \right>, & |1_+\rangle &= \left| \frac{1}{2}, +\frac{1}{2}, 1 \right>,\\
    |L_1\rangle &= \left| \frac{3}{2}, -\frac{3}{2}, 1 \right>, & |L_2\rangle &= \left| \frac{3}{2}, -\frac{1}{2}, 1 \right>,\\
    |L_3\rangle &= \left| \frac{3}{2}, +\frac{1}{2}, 1 \right>, & |L_4\rangle &= \left| \frac{3}{2}, +\frac{3}{2}, 1 \right>.
    \label{Eq:basisstates}
\end{aligned}   
\end{equation}

Here $|0_-\rangle$, $|1_-\rangle$, $|0_+\rangle$, and $|1_+\rangle$ are the computational basis states with total spin $S=1/2$ and spin projection down and up, respectively, and $|L_1\rangle$, $|L_2\rangle$, $|L_3\rangle$, and $|L_4\rangle$ correspond to the four leakage states with total spin $S=3/2$~\cite{DiVincenzo_2000}.

Since the exchange coupling preserves the total spin, the Hamiltonian can effectively be written in the $\{|0_{\pm}\rangle, |1_{\pm}\rangle\}$ basis as
\begin{align}
    H(t) = - \frac{1}{2} \left(J_{12} (t)  Z_L - \frac{J_{23} (t)}{2} \left(Z_L + \sqrt{3}  X_L \right) \right),\label{Eq:1Q}
\end{align}
where $X_L$, and $Z_L$ are the Pauli matrices. The Hamiltonian is equivalent in the two qubit subspaces $S=1/2, S_z = \pm 1/2$. In the following, the index $L$ is used for the operations in the qubit subspace and Pauli matrices without index correspond to spin operators in the normal spin basis. Note that $J_{12}$ induces rotations around the $z$-axis of the Bloch sphere and $J_{23}$ generates rotations around the vector $\vec{n}$, which lies in the $xz$-plane and has a 120\textdegree\, angle to the $z$-axis.

\subsubsection{Stabilizers and gauge operations}
For stabilizer quantum error-correcting codes, the code space is defined as the simultaneous +1 eigenspace of a set of stabilizer operators. In similar spirit, we can define the stabilizer of the EO code space as
\begin{align}
    \mathcal{S} = \mathbf{S}_{\rm tot}^2,
\end{align}
with $\mathbf{S}_{\rm tot}$ the total spin operator.
Moreover, since the spin projection $m_{{\rm tot}, z}$ acts as a gauge degree of freedom, we can even define a set of gauge operators \begin{align}
    \mathcal{G} = \{ \mathcal{S}, P_3 \},
\end{align}
where $P_3 = \cos(\frac{\theta}{2}) \id + i \sin(\frac{\theta}{2}) (c_1 XXX + c_2 YYY + c_3 ZZZ)$, with $\sum_i c_i^2 = 1$, acts on the three spins.
The logical operations can be written as 
\begin{align}
\begin{split}
    X_{\rm L} &= \frac{ie^{-i\frac{\sqrt{3}\pi}{2}}}{2} P_{\rm EO}\left[  III +\frac{1}{3} (XXI + YYI + ZZI)\right.\\
    &\left.+ \frac{1}{3}\left( 1-i\sqrt{3}e^{i\frac{\sqrt{3}\pi}{2}} \right) (IXX + IYY + IZZ)\right. \\
    &\left. + \frac{1}{3}\left( 1+i\sqrt{3}e^{i\frac{\sqrt{3}\pi}{2}} \right) (XIX + YIY + ZIZ)\right] P_{\rm EO}^{\dagger}, \label{eq:Xlog}
\end{split}\\
    Z_{\rm L} =& -\frac{1}{2} P_{\rm EO} \left( III + XXI + YYI + ZZI \right) P_{\rm EO}^{\dagger},\label{eq:Zlog}
\end{align}
where $P_{\rm EO}$ is the projector onto the EO qubit space.
If the EO qubit is operated in the subspace of a fixed projection $m_{{\rm tot}, z}=+1/2$, this corresponds to fixing the gauge degree of freedom to the $+1$ eigenstate of the gauge operator $ZZZ$.  This corresponds to $P_3$ with $\theta = \pi$, $c_1=c_2=0$, and $c_3=1$. The stabilizer then becomes
\begin{align}
    \mathcal{S} = \langle \mathbf{S}^2_{\rm tot}, ZZZ\rangle.
\end{align}

After fixing the gauge, no gauge degree of freedom is left. However, since a single spin flip on any of the three spins will change the outcome of measuring $ZZZ$, single bit flips are detectable in this scheme, whereas phase flips remain undetected. Alternatively, one can measure the eigenstate of the $XXX$ operator instead of the $ZZZ$ operator to measure any possible phase flip, and bit flips would stay undetected. 

\subsection{The singlet-only exchange-only qubit} \label{sec:methods-SOEO}
Stabilizer readout is an essential tool for quantum error detection and correction to verify whether an error has occurred. Since the EO qubit remains in the same total spin number, readout of $XXX$ or $ZZZ$ is possible in principle. However, since these two operators do not commute, they cannot be read out simultaneously. On the other hand, stabilizers with even overlap of $X$ and $Z$ commute. Thus we investigate a natural extension of the exchange-only qubit by one more spin \cite{PhysRevLett.85.1758, PhysRevB.95.241303, j1hn-k5m8}, and encode the system in the $S_{\rm tot} = 0$ subspace. Following the definition of the singlet-only exchange-only (SOEO) qubit from Ref.~\cite{j1hn-k5m8}, where the arrangement of the spins is as in Fig.~\ref{fig:SOEO-FT-QED-cicuit}a, the qubit states are given by
\begin{align}
    \ket{0} &= \ket{S_{12}} \ket{S_{34}},\\
    \ket{1} &= \frac{1}{\sqrt{3}} \left(\ket{T_{12}^{0}} \ket{T_{34}^{0}}-\ket{T_{12}^{+}} \ket{T_{34}^{-}}-\ket{T_{12}^{-}} \ket{T_{34}^{+}} \right),
\end{align}
where $\ket{S_{ij}}$ denotes the singlet state, and $\ket{T_{ij}^0}$ and $\ket{T_{ij}^\pm}$ denote the triplet states of spin pair $\{i,j\}$ with spin projection 0 and $\pm 1$.
The Hamiltonian of the system can be transformed into a basis where the two above basis states form a decoupled computational subspace which can be written as
\begin{align}
    \begin{split}
        H= -\frac{J_{12}}{2} Z_L &+ \frac{J_{13}}{4} \left( \sqrt{3} X_L + Z_L \right) \\
    &+ \frac{J_{14}}{4} \left(-\sqrt{3} X_L + Z_L \right),
    \end{split}
\end{align}
where $X_L$ and $Z_L$ act in the qubit subspace again. The logical operators of the SOEO encoding can be defined analogously to Eqs.~\eqref{eq:Xlog} and \eqref{eq:Zlog} but with an additional identity on the extra spin and changing the projector to project onto the SOEO qubit subspace. Similarly to the EO qubit encoding, $J_{12}$ induces rotations around the $z$-axis of the Bloch sphere and $J_{13}$ generates rotations around the vector $\vec{n}=-(\sqrt{3},0,1)$. Exchange pulses on $J_{14}$ enable a direct rotation around a third vector $\vec{m}=(\sqrt{3},0,-1)$. All three rotation axes lie in the $xz$-plane and enclose a 120\textdegree\, angle between each other~\cite{j1hn-k5m8}. Two of the exchange interactions, e.g.~$J_{12}$ and $J_{13}$, are sufficient to construct any rotation on the Bloch sphere, which allows for a more flexible arrangement of the four spins. This qubit encoding allows for simultaneous readout of the two stabilizers $XXXX$ and $ZZZZ$ where the four operators act on the four spins again.
With that we can perform standard stabilizer measurements that can be made fault tolerant. 

The stabilizers correspond to the ones of the smallest quantum error detecting (QED) code, the [[4,2,2]] code, which encodes two logical qubits into four physical qubits with code distance two~\cite{vaidman1996error} (see App.~\ref{App:4-2-2-code}). By reading out the $XXXX$ and $ZZZZ$ stabilizers, the system is projected onto the same subsystem code space. However, the logical operators within the subspace differ from the standard logical operators in the [[4,2,2]] code. In fact, in our case, only one logical qubit is encoded, while the remaining degree of freedom acts as a gauge degree of freedom that is not used further. We evaluate the code capacity of the encoding using perfect $XXXX$ and $ZZZZ$ stabilizer readout in App.~\ref{App:CodeCapacity} and find a reduction in the qubit error rate for both noise models considered.

\subsubsection{Stabilizer readout and fault tolerance}

\begin{figure}[ht]
    \centering
    \includegraphics[width=0.99\linewidth]{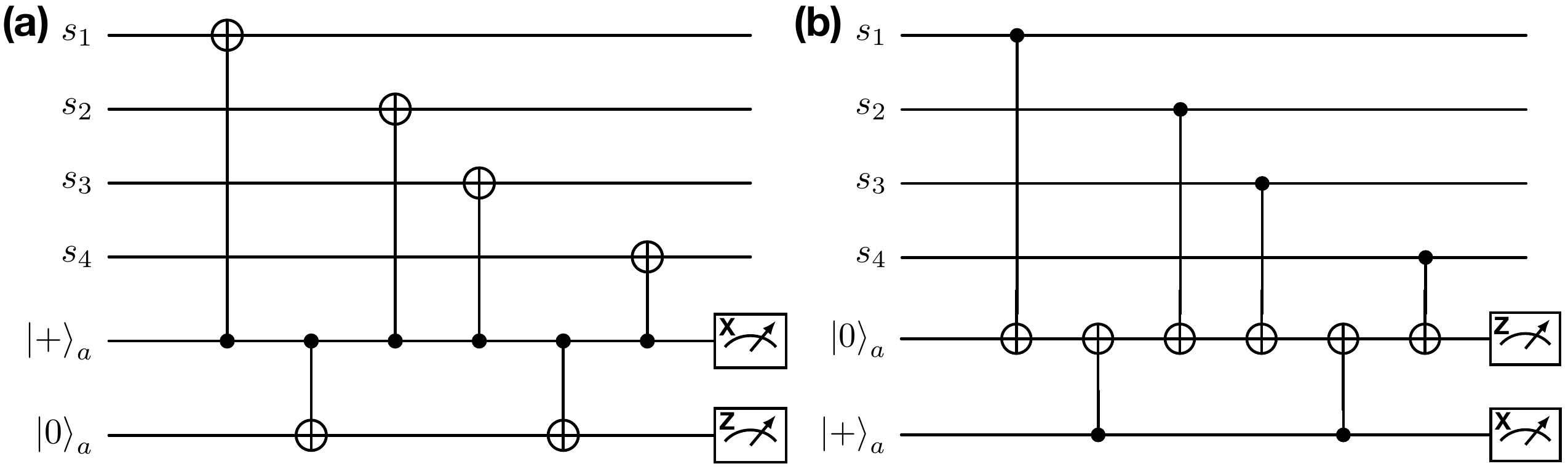}
    \caption{
    Fault-tolerant $XXXX$ and $ZZZZ$ measurements using flags.}
    \label{fig:FTStabilizerMeasurement}
\end{figure}

The $XXXX$ and $ZZZZ$ stabilizers of the SOEO qubit can be read out using an auxiliary spin and four CNOT gates. However, stabilizer readout can introduce additional errors to the encoded state due to errors in the state preparation of the auxiliary spin, during quantum operations, or in the measurement of the auxiliary spin. It is thus important to handle these errors in a way that those arising from the stabilizer measurement gadget are at least detectable by a noiseless round of error detection \footnote{This last step is a task left to the decoder of the measured qubit state afterwards. We note that the SOEO qubit is usually read out  using a two-body measurement and thus is not fault tolerant. Therefore, it requires a decoding strategy that preserves fault tolerance.}. This can be done using flag qubits \cite{Chao_2018, Chamberland_2018, Chao_2020}. 
Fault-tolerant qubit state encoding can be realized using a non-fault-tolerant encoding circuit followed by a verification step~\cite{Linke_2017, Chamberland_2019, Gupta_2024, PRXQuantum.6.020330}. We will briefly discuss the fault-tolerant encoding and decoding for SOEO qubit in a later section. 
Let us assume that the SOEO qubit has been prepared fault tolerantly, such that at most a single spin error remains. We will refer to this as a weight-one fault. Our goal is to increase the qubit's lifetime and so preserve the quantum memory. Our approach defines the SOEO qubit encoding as an error detection code, which is described by the same stabilizers as the ones of the $[[4,2,2]]$ subsystem code summarized in App.~\ref{App:4-2-2-code}. Figure~\ref{fig:FTStabilizerMeasurement} shows the fault-tolerant $XXXX$ and $ZZZZ$ stabilizer readout circuits. Together they detect any weight-one fault but also introduce at most a weight-one fault if a single component in the circuit fails, which is again detectable by the code. The first four lines typically represent the data qubits which would correspond to the four spins of the SOEO qubit in our case. The lower two lines are LD spin qubits which represent the auxiliary (upper) and the flag qubit (lower). The measurement outcome of the auxiliary qubits is called the syndrome. 
In case no error has occurred, the syndrome is trivial and the encoded spins are in the $+1$ eigenstate of the code space defined by the stabilizers. If a single $Z$ fault on a single spin has happened, the measurement outcome in (a) will flip. On the other hand, a single $X$ fault will flip the measurement outcome of (b). This way any combination of single $X$ and single $Z$ errors can be detected. If some component during the stabilizer measurement circuit fails, it either results in a weight-one fault which is detectable by the code, or it creates a higher-weight error that is then detected by the flag qubit. Only if both, auxiliary and flag outcomes are trivial, we accept the run and proceed with the quantum computation. This postselection gives us a non-deterministic way of increasing the qubit performance. 

\begin{figure}[ht]
    \centering
    \includegraphics[width=0.98\linewidth]{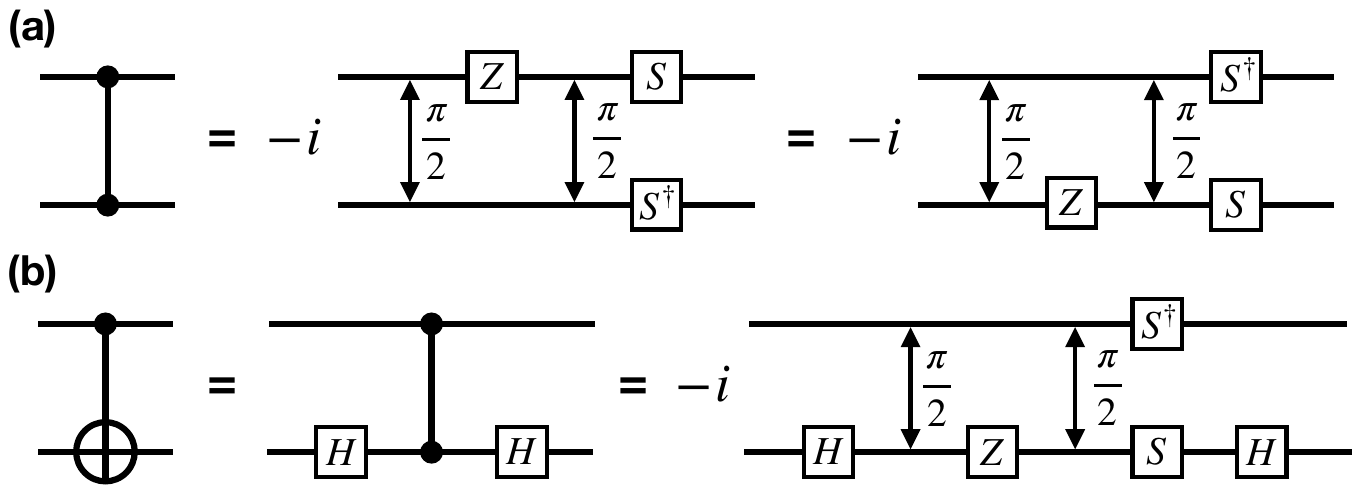}
    \caption{CZ and CNOT implementations between two spins.}
    \label{fig:CNOTbetweenSpins}
\end{figure}

In total, to realize the fault-tolerant stabilizer readout from Fig.~\ref{fig:FTStabilizerMeasurement}, we use circuits in Fig.~\ref{fig:CNOTbetweenSpins}, where realizations of the CZ and CNOT gates are shown. The double arrows depict $\sqrt{\rm SWAP}$ operations which correspond to $\pi/2$ pulses of the exchange interaction between the two spins. 
One can construct the CNOT sequence such that single-spin operations are required on only one of the spins. As a matter of fact, the $S^{\dagger}$ operation on the upper spin in Fig.~\ref{fig:CNOTbetweenSpins} can in principle be shifted to the lower spin using SWAP operations. Nevertheless, for SOEO qubit a $S^{\dagger}$ operation on all four spins is also a stabilizer operation of the state and thus does not need to be physically executed after all. The stabilizer readout thus results in the circuits in Figs.~\ref{fig:SOEO-FT-QED-cicuit}c and d.
Note that control and target qubits of the CNOT operation can be swapped using additional SWAP pulses. 

To evaluate the fault tolerance of the constructed QED cycle in Section~\ref{sec:results-SOEO}, we conceptually append a noiseless error-detection cycle after the noisy implementation. The first cycle is subject to circuit-level noise, while the second is assumed to be ideal and serves only to determine whether the noisy cycle has produced an undetected error. If the ideal cycle reports no syndrome, the encoded state is accepted and compared to the ideal code state. A fault-tolerant QED cycle should ensure that any error arising from a single physical fault is detected by the subsequent ideal cycle, so that undetected failures occur only due to multiple faults. Consequently, the conditional probability of an undetected logical error is expected to scale quadratically with the physical spin error probability \cite{aliferis2005quantumaccuracythresholdconcatenated}. While the ideal error-detection cycle serves as a numerical diagnostic, the corresponding experimental verification can be performed by a final fault-tolerant measurement of the encoded state followed by classical decoding \cite{Postler_2022, Egan2021, Krinner_2022,  Bluvstein_2023}.

\subsubsection{Encoding and decoding}
\begin{figure}[ht]
    \centering
    \includegraphics[width=0.55\linewidth]{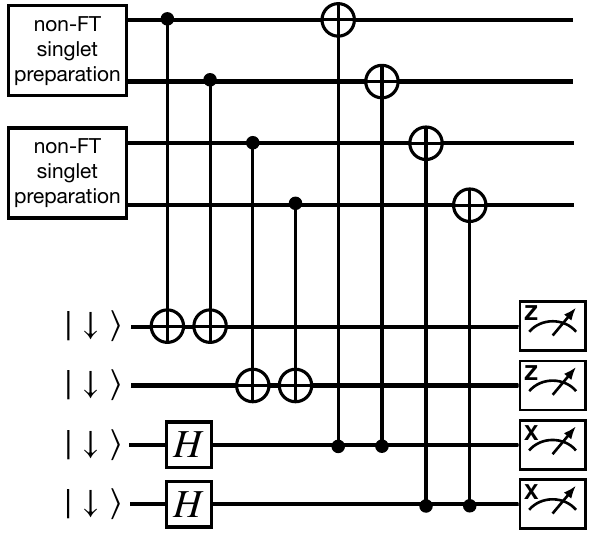}
    \caption{Fault-tolerant encoding of the SOEO qubit using four auxiliary spins.}
    \label{fig:FTencoding}
\end{figure}

We have assumed that after initialization there is only a single spin fault on the state. In practice, initialization requires preparing two singlet states followed by appropriate exchange pulses. Singlet pairs typically involve two-spin operations. The electrons start in one quantum dot, in which the state relaxes to the ground state, the spin singlet. Then, one electron is pulsed into the neighboring quantum dot. This initialization can be easily made fault tolerant. Since the SOEO $\ket{0}$ state is an eigenstate of $Z_L$, one can simply measure the operators $XXII$, $IIXX$, $ZZII$, and  $IIZZ$ to detect any harmful error, as shown in Fig.~\ref{fig:FTencoding}. Each measurement requires one auxiliary spin and two CNOT operations. Measurements of weight-two operators are fault tolerant on their own and do not require any flags to detect logical errors. 

In stabilizer quantum codes, the final code state at the end of an algorithm can be obtained using transversal qubit measurements, i.e., each qubit is measured separately, and is combined with classical decoding to obtain the most probable code state. The exchange-only qubit is typically read out using singlet-triplet measurements on spins 1 and 2, which is not fault tolerant, as it can introduce weight-two errors. A natural way to read out the singlet-only exchange-only qubit is to do singlet-triplet measurements on spins 1 and 2 and also on 3 and 4. Although this readout scheme is also not fault tolerant, it can already detect some errors by checking whether both measurements yield the same result. If this is not the case, an error must have happened and the run can be discarded. Nevertheless, to ensure quadratic scaling for the logical error rate, a fault-tolerant qubit readout scheme is desirable. Similar to the encoding, fault-tolerant measurements of the operators $XXII$, $IIXX$, $ZZII$, and  $IIZZ$ could be used to decode the information.  

\subsection{Error model} \label{sec:error-model}
To estimate the performance of the proposed protocols of this paper, we have considered different noise models. In Fig.~\ref{fig:JJJgate}, we have assumed fluctuations in the exchange interactions, which we expect to be the main error source of such a finely calibrated gate suffering from charge noise. For our simulations, we use single-spin depolarizing noise 
\begin{equation}
    \mathcal{E}(\rho) = (1-p) \rho + \frac{p}{3} \sum_{k=1}^{3} P_k^{s_i} \rho P_k^{s_i} . \label{eq:error-on-single-qubit-gates}
\end{equation}
Here, $P_0^{s_i} = \mathbb{1}$, $P_1^{s_i} = X$, $P_2^{s_i} = Y$, and $P_3^{s_i} = Z$ are the identity and the Pauli matrices acting on spin $s_i$. No error occurs on spin $s_i$ with probability $(1-p)$ and a single Pauli error occurs on a spin $s_i$ with probability $\frac{p}{3}$ each. For idling noise, we take $p=p_{\rm idle}(t)$ which is derived later in this section. 

For the simulations of the fault-tolerant stabilizer readout in Fig.~\ref{fig:SOEO-QED-simulation} we have used circuit-level noise. Noiseless preparation is followed by a depolarizing noise channel [Eq.~\eqref{eq:error-on-single-qubit-gates}] with probability $p_{\rm spin}$ on each of the six spins individually. Each gate used in the stabilizer readout circuit in Figs.~\ref{fig:SOEO-FT-QED-cicuit}c and ~\ref{fig:SOEO-FT-QED-cicuit}d is likewise followed by a depolarizing noise channel with error probability $p_{\rm spin}$ on the spins it acts on. For single-spin gates, this corresponds to Eq.~\eqref{eq:error-on-single-qubit-gates} with $p=p_{\rm spin}$ (also see Fig.~\ref{fig:SOEO-FT-QED-cicuit}b.ii). For a two-qubit gate acting on spins $s_i$ and $s_j$, the error channel corresponds to
\begin{equation}
    \mathcal{E}(\rho) = (1-p) \rho + \frac{p}{15}\sum_{\{k, l\}\neq \{0,0\}}  (P_k^{s_i} \otimes P_l^{s_j}) \rho (P_k^{s_i} \otimes P_l^{s_j}).\label{eq:error-on-two-qubit-gates}
\end{equation}
Here, no error occurs with probability $(1-p)$ and any of the two-spin Pauli operators $P_k^{s_i} \otimes P^{s_j}_l$, with $k,l\in\{0,\ldots, 3\}, \{k,l\}\neq\{0,0\}$, acting on spins $s_i$ and $s_j$, occurs with probability $p/15$ each. The two-spin circuit-level noise is depicted in Fig.~\ref{fig:SOEO-FT-QED-cicuit}b.i. In the memory experiment (Section~\ref{sec:memory}), additionally all idling spins undergo single-spin depolarizing noise with $p=p_{\rm idle}(t)$.

\subsubsection{Effective idling noise models}\label{sec:idling_noise}
We now derive the two types of idling noise models, being single-qubit depolarizing noise with 
\begin{align}
p=p_{\mrm{idle}}(t)&=\frac{3}{4}\left(1-e^{-(t/T)^\zeta}\right),\label{eq:p_idle}
\end{align}
where $\zeta=2$ for type-(i) decay of $p_{\rm idle}(t)$ and $\zeta=1$ for type-(ii) decay. 

A major noise source in EO qubits is formed by stray substrate nuclei with a nonzero magnetic moment \cite{kerckhoff2021magnetic}. As in Ref. \cite{kerckhoff2021magnetic}, for a single spin, we may describe this magnetic environment semiclassically as 
\begin{equation}\label{eq:hamiltonian}
H(t)=-g\mu_B\bB(t)\cdot \mathbf S,
\end{equation}
with $g$ the electron's $g$-factor and $\mu_B$ the Bohr magneton. The $\bB_i(t)$ are stochastic real variables describing the magnetic environment, and $\mathbf S$ is the spin vector $\mathbf S_i=\sigma_i/2$ (in units where $\hbar=1$), where $\sigma_i$ ($i\in\{1,2,3\}$) are the Pauli matrices. Thus, $H(t)$ is of the form 
\begin{equation}\label{eq:expansion}
H(t)=\sum_i b_i(t) \sigma_i,    
\end{equation}
with $b_i(t)=-g\mu_B B_i(t)/2$. 

We partially follow Ref.~\cite{hangleiter2021filter,green2013arbitrary} to obtain the effective noise models. The time-dependent Hamiltonian generates the quantum channel $\mc U(t)=U(t)(\,\cdot\,)U^\dagger(t)$, with $U(t)$ the standard time-evolution operator. The quantum channel $\mc U(t)$ satisfies the Liouville-von Neumann equation
\begin{align}
    \frac{\mrm d \mc U(t)}{\mrm d t}&= -i\mc L(t)\mc U(t), && \mc L(t)=[H(t),\,\cdot\,].
\end{align}
This is solved by the Magnus expansion
\begin{equation}
    \mc U(t)=e^{-it \sum_{m=1}^\infty \mc L_m(t)},
\end{equation}
with $\mc L_m(t)$ the $m$th term in the Magnus expansion. 
The first two terms are 
\begin{align}
    \mc L_1(t)&=\frac{1}{t}\int_0^t \mrm d\tau\, \mc L(\tau),\\
    \mc L_2(t)&=-\frac{i}{2t}\int_0^t\mrm d\tau_1\int_0^{\tau_1}\mrm d\tau_2\, [\mc L(\tau_1),\mc L(\tau_2)].
\end{align}
So far, we have shown how to obtain the channel $\mc U(t)$ given a realization of the noisy field $\bB(t)$. We denote the average over these realizations by $\langle\,\cdot\,\rangle$. The cumulant $\mc K(t)$ is defined via
\begin{equation}
    \langle \mc U(t) \rangle=e^{\mc K(t)}. 
\end{equation}
For zero-mean Gaussian noise, up to fourth order in the noise strength~\cite{hangleiter2024erratum},
\begin{equation}
    \mc K(t)\approx-it \langle \mc L_2(t)\rangle-\frac{t^2}{2}\langle \mc L_1^2(t)\rangle.
\end{equation}
Higher-order terms generally do not vanish,  even when there is no deterministic part in $H(t)$, due to the noncommutativity of $H(t)$ at different times.

Using the expressions for the first two terms in the Magnus expansion, and the expansion of the Hamiltonian in the Pauli basis [Eq. \eqref{eq:expansion}], one obtains
\begin{align}
\langle \mc L_2(t)\rangle&=-\frac{i}{2t}\sum_{i,j=1}^3\int_0^t\!\!\mrm d\tau_1\!\!\int_0^{\tau_1}\!\!\mrm d\tau_2\, \langle b_i(\tau_1)b_j(\tau_2)\rangle \big[[\sigma_i,\sigma_j],\,\cdot\,\big],\\
\langle \mc L_1^2(t)\rangle&=\frac{1}{t^2}\sum_{i,j=1}^3\!\int_0^t\mrm d\tau_1\int_0^{t}\!\!\mrm d\tau_2\, \langle b_i(\tau_1)b_j(\tau_2)\rangle \big[\sigma_i,[\sigma_j,\,\cdot\,]\big].
\end{align}
We now assume that the noise in the $x,y,z$ directions is identical and independently distributed, meaning that $\langle b_i(\tau_1)b_j(\tau_2)\rangle\propto\delta_{ij}$ and that $\langle b_i(\tau_1)b_i(\tau_2)\rangle\equiv C(\tau_1,\tau_2)$ is independent of $i$. This immediately gives $\langle \mc L_2(t)\rangle=0$.

The single-spin Pauli transfer matrix $T$ of a superoperator $\La$ is defined as $[T(\La)]_{kl}=\frac{1}{2}\mrm{tr}[\sigma_k\La(\sigma_l)]$, with $k,l\in\{0,1,2,3\}$, where $\sigma_0:=\id$.
Now consider the transfer matrix $T$ of $\langle \mc L_1^2(t)\rangle$. Using the commutation relations for Pauli operators and their trace orthogonality, we have 
\begin{equation}
\sum_i\tr\left(\sigma_k \big[\sigma_i,[\sigma_i,\sigma_l]\big]\right)=\left\{
\begin{array}{cl}
     0&  (k=0\vee l=0 )\\
     16\delta_{kl} & (\text{otherwise}).
\end{array}
\right.
\end{equation} 
With this, we have for $k,l\neq 0$
\begin{equation}
[T(\langle \mc L_1^2(t)\rangle)]_{kl}=\frac{8\delta_{kl}}{t^2}\int_0^t\mrm d\tau_1\int_0^{t}\mrm d\tau_2\, C(\tau_1,\tau_2),
\end{equation}
where all other entries vanish. 

With the above, 
\begin{align}
T(\langle \mc U(t)\rangle)&=e^{T(\mc K(t))}\\
&=e^{-t^2 T(\langle \mc L_1^2(t)\rangle)/2}\\
&=\mrm{diag}(1,e^{-4 I(t)},e^{-4 I(t)},e^{-4 I(t)}),
\end{align}
where
\begin{align}
I(t)&=\int_0^t\mrm d\tau_1\int_0^{t}\mrm d\tau_2\, C(\tau_1,\tau_2).
\end{align}

A quantum channel $\La$ with a diagonal transfer matrix is a Pauli channel $\La(\,\cdot\,)=\sum_{i=0}^3 p_i \sigma_i(\,\cdot\,)\sigma_i$, where $p_i$ is a probability distribution. The $p_i$ are related to the diagonal entries $f_i$ of the transfer matrix by a Walsh-Hadamard transformation, $p_i=\frac{1}{4}\sum_{j}(-1)^{\langle i,j \rangle} f_j$, where $\langle i,j \rangle=0$ if $\sigma_i$ commutes with $\sigma_j$, and $\langle i,j \rangle=1$ otherwise. This gives
\begin{align}
p_0(t)&=\frac{1}{4}\left(1+3e^{-4 I(t)}\right), \\ p_{1}(t)&=p_{2}(t)=p_{3}(t)=\frac{1}{4}\left(1-e^{-4I(t)}\right).
\end{align}
Within the cumulant approximation above, the ensemble-averaged noise process described by Eq.~\eqref{eq:hamiltonian} is therefore described by an effective depolarizing channel with
\begin{equation}
    p_{\mrm{idle}}(t)=\frac{3}{4}\left(1-e^{-4I(t)}\right).
\end{equation}

We now consider different options for $I(t)$. Assuming the correlator $C(\tau_1,\tau_2)$ depends only on the time difference, we may define the two-sided spectral density $S(\omega)$ by $C(\tau_1,\tau_2)=\int_{-\infty}^\infty \frac{\mathrm d\omega}{2\pi} S(\omega)e^{-i\omega(\tau_1-\tau_2)}$. We then have
\begin{equation}
  I(t)=\int_{-\infty}^\infty \frac{\mathrm d\omega}{2\pi} F(\omega,t)S(\omega),
\end{equation}   
with the filter function $F(\omega,t)=4 \sin^2(\omega t/2)/\omega^2$. We carry out the integral choosing quasistatic, white, and 1/$f$ noise spectra~\cite{ithier2005decoherence}, and report the results in Table~\ref{tab:spectra}. The $1/f$ noise power spectrum was experimentally identified as a realistic spectrum for EO qubits with a noise model of Eq.~\eqref{eq:hamiltonian} \cite{kerckhoff2021magnetic}. This spectrum requires an infrared cutoff $\omega_{\mathrm{ir}}$ to converge. We choose $S(\omega)=\Theta(\lvert \omega \rvert - \omega_\mathrm{ir})2\pi S_{1/f}/\lvert \omega \rvert$, with $\Theta$ the Heaviside step function. 

\begin{table*}[ht]
\centering
\renewcommand{\arraystretch}{1.6}
\setlength{\tabcolsep}{12pt}
\begin{tabular}{|c|c|c|}
\hline
$S(\omega)$ & $I(t)$ & Noise description \\
\hline
$2\pi S_{\mathrm{qs}}^2 \delta(\omega)$
&
$S_{\mathrm{qs}}^2 t^2$
&
quasistatic
\\
$S_{\mathrm{w}}$
&
$S_{\mathrm{w}}t$
&
white
\\
$\Theta\!\left(\lvert \omega\rvert-\omega_{\mathrm{ir}}\right)
\dfrac{2\pi S_{1/f}}{\lvert \omega\rvert}$
&
$2S_{1/f}t^2
\left[
-\ln\!\left(\omega_{\mathrm{ir}}t\right)
+ O(1)
\right]$
&
$1/f$
\\
\hline
\end{tabular}
\caption{Noise spectra and corresponding exponents $I(t)$ for $p_{\rm idle}(t)$. }
\label{tab:spectra}
\end{table*}

The quasistatic noise spectrum gives type-(i) decay of $p_{\rm idle}(t)$, whereas the white noise spectrum gives type-(ii) decay. The $1/f$ noise spectrum gives type-(i) decay with a logarithmic correction. To see the effect of this correction, note that $-t^2\ln(\omega_{\mathrm{ir}}t)$ has slope $2+1/\ln(\omega_{\mathrm{ir}}t)$ on a log-log plot, meaning that the effective exponent of $t$ is well approximated by 2 for $\omega_{\mathrm{ir}}t\ll 1$. 

\subsection{Combining idling noise with circuit-level noise}
Circuit-level noise is a rudimentary noise model. We extend it by incorporating idling noise at a comparable level of approximation. The resulting model is not intended to provide a fully microscopic description of the physical noise processes, but neither is the underlying circuit-level model. It nevertheless captures the qualitative distinction between errors accumulated during idle periods and errors associated with circuit operations.

Our idling model admits a division into temporal channel segments because the corresponding family of reduced channels is completely positive (CP)-divisible. (Under suitable regularity conditions, this means that it admits a description by a time-local Gorini–Kossakowski–Lindblad–Sudarshan generator~\cite{rivas2014quantum}.) A two-parameter family of channels $\{\mc A_{t_b,t_a}\mid t_b\geq t_a\geq  0\}$ is called CP-divisible if, for all intermediate times $t$ $(t_b \geq t\geq t_a)$, $\mc A_{t_b,t_a}=\mc A_{t_b,t} \circ \mc A_{t,t_a}$~\cite{rivas2014quantum}. Define the channels $\mc E_{t_b,t_a}$ as depolarizing channels [Eq.~\eqref{eq:error-on-single-qubit-gates}], with depolarizing probability
\begin{equation}\label{eq:pab}
p_{t_b,t_a}=\frac{3}{4}\left(1-e^{-\left(\frac{t_b}{T}\right)^\zeta+\left(\frac{t_a}{T}\right)^\zeta}\right).
\end{equation}
The single-spin depolarizing channel [Eq.~\eqref{eq:error-on-single-qubit-gates}] can be rewritten as
\begin{equation}\label{eq:aternative_depol}
\mathcal E(\rho)=(1-\frac{4}{3}p)\rho+\frac{4}{3}p\frac{\id}{2},
\end{equation}
without redefining $p$. With $p=p_{t_b,t_a}$ and $(1-4p/3)=\exp[-\left(t_b/T\right)^\zeta+\left(t_a/T\right)^\zeta]$, it is straightforward to see that $\mc E_{t_b,t_a}=\mc E_{t_b,t} \circ \mc E_{t,t_a}$, and hence the two-parameter family of idling channels $\{\mc E_{t_b,t_a}\mid t_b\geq t_a\geq  0\}$ is CP-divisible. Furthermore, note that $p_{t,0}=p_{\rm idle}(t)$ [see Eq.~\eqref{eq:p_idle}] for all $t\geq 0$, showing that $\mc E_{t,0}$ is the correct depolarizing channel. That is, we can divide the depolarizing channel $\mc E_{t_f,0}$ at will
\begin{equation}
\mathcal E_{t_f,0}=\mathcal E_{t_f,t_{f-1}} \circ \ldots \circ \mathcal E_{t_2,t_1} \circ \mathcal E_{t_1,0}.
\end{equation}
Although the family of idling channels is CP-divisible, it is not Markovian in the sense of the semigroup property~\cite{rivas2014quantum}, unless $\zeta=1$.

\begin{figure}[h]
    \centering
    \includegraphics[width=0.99\linewidth]{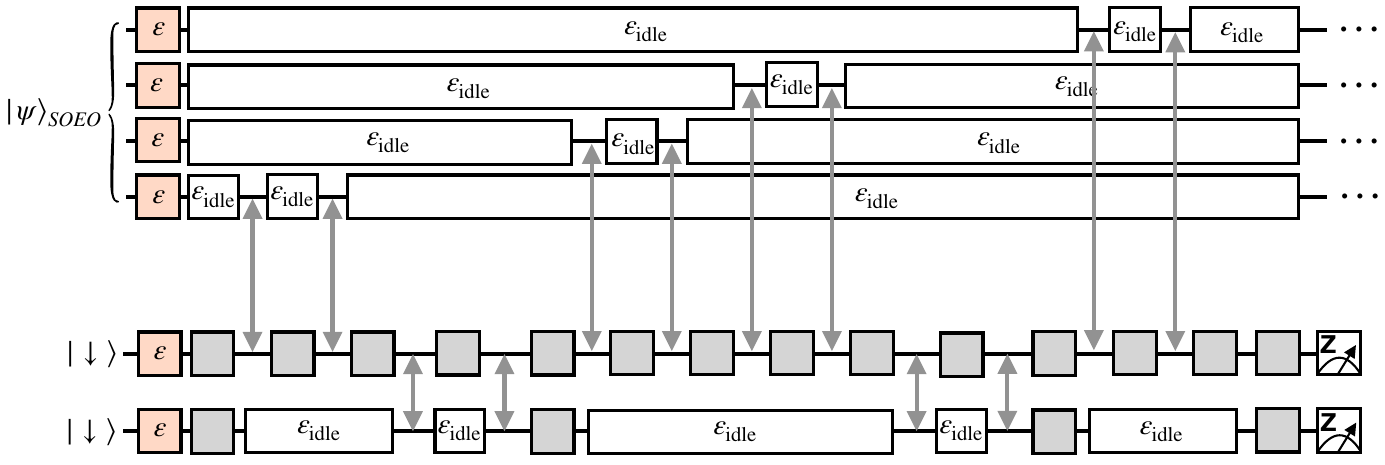}
    \caption{Simulation circuit depicting the noise modeling in the memory experiment. Red boxes correspond to initialization errors, gray boxes and double arrows correspond to noisy quantum gates modeled as in Fig.~\ref{fig:SOEO-FT-QED-cicuit}b. White boxes correspond to idling noise, where $\mc E_{\rm idle}$ corresponds to $\mathcal E_{t_b,t_a}$ in the main text, with the appropriate start time $t_a$ and end time $t_b$. The auxiliary and flag spins are reset after measurement.}
    \label{fig:idlingNoise}
\end{figure}

We use divisibility to define a circuit-level approximation in which the intermediate maps $\mathcal E_{t_b,t_a}$ are treated as the idling-noise channels acting between circuit operations. During the gate interval, we replace the idling channel with the noisy gate channel, rather than applying gate noise in addition to idling noise. Consider a qubit undergoing idling noise from time $t_a$ to $t_b$. The idling period is divided at times $t$ and $t+t_g$, giving the decomposition $\mathcal E_{t_b,t_a}= \mathcal E_{t_b,t+t_g} \circ \mathcal E_{t+t_g,t} \circ \mc E_{t,t_a}$. A single- or multi-qubit gate $G$ with unitary channel $\mc G$ and noise channel $\mc E$ is inserted beginning at time $t$ and lasting for a time $t_g$. This is modeled as
\begin{equation}
\tilde{\mathcal E}_{t_b,t_a}
=
\mathcal E_{t_b,t+t_g} \circ (\mc E \circ \mathcal G) \circ \mc E_{t,t_a}.
\end{equation}
The application of this error model to the syndrome extraction circuits is depicted schematically in Fig.~\ref{fig:idlingNoise}. 

We emphasize that this is an effective channel description of the idling noise. For example, for quasistatic noise it reproduces the characteristic quadratic short-time decay, but does not retain all temporal correlations of the underlying quasistatic stochastic process. Also, Eq.~\eqref{eq:pab} shows that, for type-(i) noise, the idling error accumulated over a fixed interval from $t$ to $t+t_g$ increases with the absolute time $t$. This would imply a time-dependent contribution to the physical gate error. In practice, however, randomized benchmarking typically characterizes gate performance through a single effective error rate averaged over the gates and experimental sequences~\cite{magesan2012efficient}. We therefore model the gate error by a time-independent effective depolarizing channel with error probability $p_{\rm spin}$, which should be interpreted as representing such an averaged gate error.

\subsection{The quantum Zeno effect}\label{sec:zeno}

The quantum Zeno effect states that the dynamics of a quantum system can be completely frozen by continuous observation \cite{misra1977zenos}. It has also been considered in the case of frequent, but not continuous observation, which likewise causes a slowdown of the dynamics \cite{itano1990quantum}, given that observation is frequent enough. Already in the first papers on quantum error-detecting codes \cite{zurek1984reversibility, vaidman1996error}, the quantum Zeno effect was identified as a mechanism through which quantum error detection can prevent errors from happening in the first place.

The quantum Zeno effect is usually treated in a scenario where the free evolution of the quantum system is unitary \cite{misra1977zenos,itano1990quantum,zurek1984reversibility, vaidman1996error,peres1980zeno}, where a key observation is that, upon measurement, the probability of being in the initial state decays with $O(t^2)$ instead of $O(t)$ \cite{peres1980zeno}. Here, we treat the initial time evolution when the free evolution is described by a time-dependent depolarizing channel, and specifically for type-(i) and type-(ii) decay of $p_{\rm idle}(t)$. Consider a spin initialized in some state $\rho=\ket \psi\!\bra{\psi}$ undergoing a depolarizing channel [Eq.~\eqref{eq:error-on-single-qubit-gates}] with error probability $p(t)$. After the channel, a projective measurement $\{\rho, \id-\rho\}$ is performed that interrogates if the system is still in state $\rho$. Using the alternative form of the depolarizing channel [Eq.~\eqref{eq:aternative_depol}], the probability that the measurement projects the state $\mathcal E_t[\rho]$ back to $\rho$ is 
\begin{equation}\label{eq:survival}
P_t(\ket\psi)=\bra{\psi}\mathcal E_t(\rho) \ket\psi=1-\frac{2}{3}p(t). 
\end{equation}

In the standard setting of unitary dynamics, the Zeno time $\tau_Z$ is obtained from the leading nontrivial term in the short-time expansion of the probability that the state is projected back to $\ket{\psi}$~\cite{facchi1998temporal}. This gives $\tau_Z=1/\Delta H$, where $(\Delta H)^2$ is the variance of the energy in the initial state. Applying the same definition to our setting, with $p(t)=p_{\rm idle}(t)$ as in Eq.~\eqref{eq:p_idle}, yields $\tau_Z=2^{1/\zeta}T$. This is consistent with the role of the relevant dynamical timescales, set by $1/\Delta H$ and $T$ respectively. One QED cycle takes $t_c=81t_g$ (Fig.~\ref{fig:SOEO-FT-QED-cicuit}, using that $\pi$-pulses take up two gate times). With the effective coherence times $T_{\mathrm{eff}}=\tau$ of Section~\ref{sec:memory}, we have $t_c/T_{\mathrm{eff}} \approx 0.1$ for type-(i) and $t_c/T_{\mathrm{eff}}\approx 0.07$ for type-(ii) noise. Thus, the time $t_c$ between measurements is below the quantum Zeno time, and we may expand Eq.~\eqref{eq:survival} for $t=t_c$, yielding
\begin{align}
 P_{t_c}(\ket\psi)=1-\frac{1}{2}  \left (\frac{t_c}{T}\right)^\zeta+O\!\left[\left (\frac{t_c}{T}\right)^{2\zeta}\right]. 
\end{align}
That is, it is only for type-(i) noise that we observe a quadratic decay of $P_{t_c}(\ket\psi)$ that is similar to the quadratic decay causing the quantum Zeno effect in the unitary case. 

\section*{Author contributions}
I.H. conceived the study and developed the protocols. I.H. and M.S. developed the quantum error-correction protocols, analyzed the error mechanisms, and performed the numerical simulations.
J.K. contributed to the implementation of the numerical simulations and derived the idling noise models. All authors discussed the results and contributed to the final version of the manuscript.

\section*{Acknowledgments} 
We thank Katrin Bolsmann, Josias Old, Luis Colmenarez, and Markus Müller for insightful discussions and valuable comments on quantum error correction and fault tolerance.
I.H. acknowledges support by  European Union’s Horizon Europe research and innovation programme under Grant Agreement No. 101114305 (“MILLENION-SGA1” EU Project).
J.K. acknowledges support by the Munich Quantum Valley (K-8), which is supported by the Bavarian state government
with funds from the Hightech Agenda Bayern Plus.

\appendix

\section{Gauge projection for the exchange-only qubit}\label{App:JJJ-map}
We derive the linear map for the gauge projection of Section~\ref{sec:JJJgate}. We start with the product state, $\rho \otimes \ketbra{\downarrow}{\downarrow}$, of the EO qubit in state $\rho$ and the LD qubit initialized in spin-down. The operation $U(t_g)$ is applied to the system with $t_g = \pi/J$. The time evolution is given by Eq.~\eqref{eq:UJJJ-gate}.
A subsequent measurement of the LD qubit in the $z$-basis leads to the following map
\begin{align}
    \rho \rightarrow \Tr_{\rm LD}[U(t_g) (\rho \otimes \ketbra{\downarrow}{\downarrow} ) U^{\dagger}(t_g)].
\end{align}
Here, $\Tr_{\rm LD}[\cdot]$ denotes the partial trace over the LD spin qubit. We write 
\begin{align}
    \begin{split}
    \Tr_{\rm LD}[U(t_g) (\rho \otimes \ketbra{\downarrow}{\downarrow} ) U^{\dagger}(t_g)] =  A_0 \rho A_0^{\dagger}
    + A_1 \rho A_1^{\dagger},
    \end{split}
\end{align}
where $A_{0} = \bra{\downarrow}U(t_g)\ket{\downarrow}$ and $A_{1} = \bra{\uparrow}U(t_g)\ket{\downarrow}$. Evaluating the matrix elements yields
\begin{align}
    A_{0} &= \ketbra{0_-}{0_-} + \ketbra{1_-}{1_-} 
    + \sum_{i=1}^{4} \ketbra{L_i}{L_i} ,\\
    A_{1} &= i \ketbra{0_-}{0_+} + i \ketbra{1_-}{1_+} .
\end{align}

\section{The smallest error detection code} \label{App:4-2-2-code}
The smallest error-detecting code, the [[4,2,2]] code, encodes two logical qubits into four physical qubits with code distance two \cite{vaidman1996error}, i.e., it can detect weight-one errors. The stabilizer group of the code is
\begin{align}
    \mathcal{S} = \langle XXXX, ZZZZ \rangle.
\end{align}
The logical operators of the two qubits are given by
\begin{align}
    &X_L^{1} = XXII,   &&Z_L^{1} = ZIZI,\\
    &X_L^{2} = XIXI,   &&Z_L^{2} = ZZII.
\end{align}
They all commute with the stabilizers as they have even overlapping support. The logical operators of the first qubit commute with the ones of the second qubit, and logical operators of one qubit anti-commute with each other. Using only one of the qubits for encoding quantum information results in a gauge degree of freedom, similar to the case of the SOEO qubit.  

\section{SOEO code capacity} \label{App:CodeCapacity}
\begin{figure}[h]
    \centering
    \includegraphics[width=0.99\linewidth]{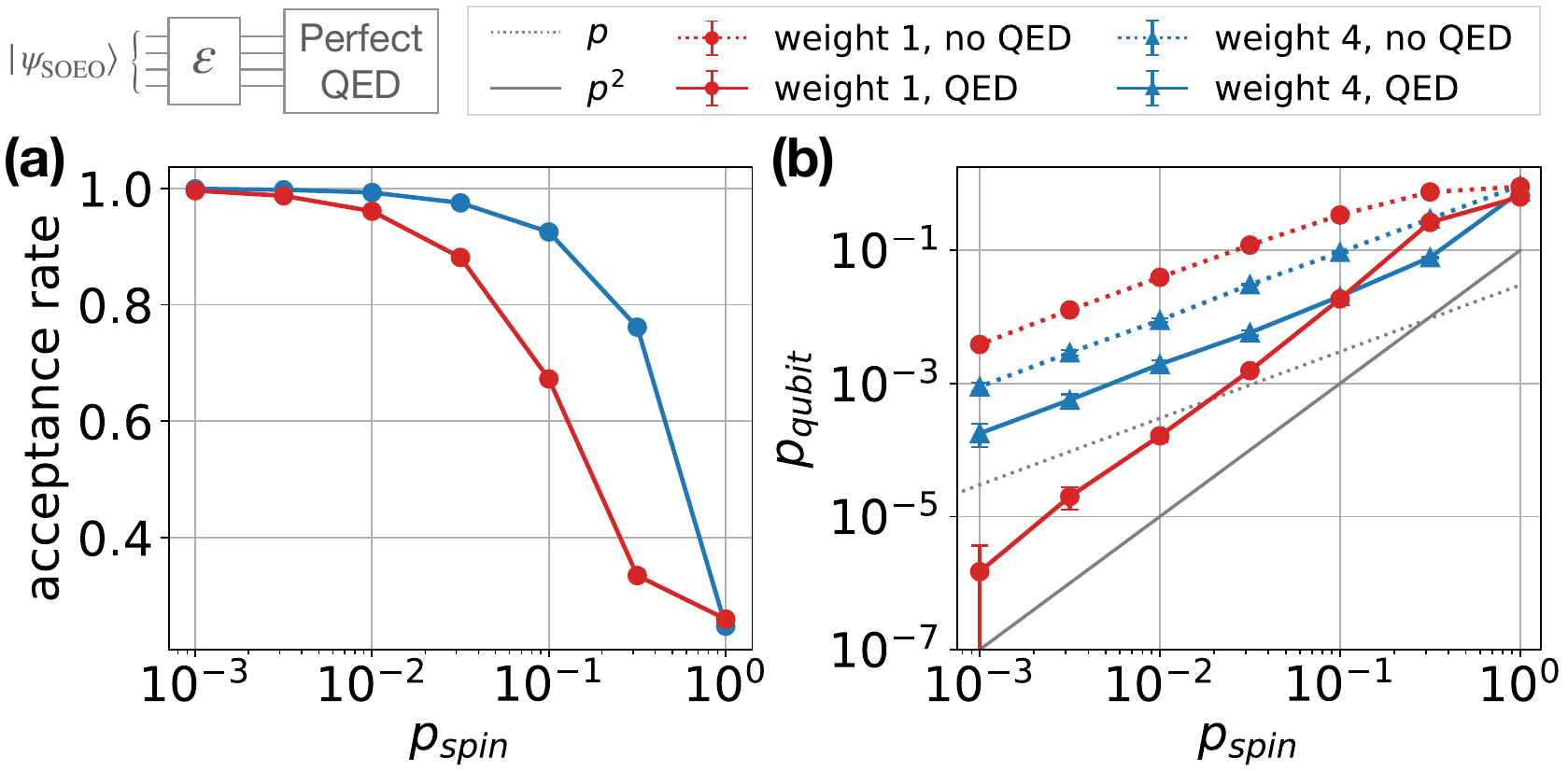}
    \caption{Simulations of the code capacity. Starting from a perfect SOEO state followed by depolarizing noise $\mathcal{E}$, a noise-free round of quantum error detection is performed. (a) Acceptance rate and (b) qubit error probability $p_{\rm qubit}$ depending on the spin error probability $p_{\rm spin}$ for the single-spin (red) and four-spin depolarizing noise channel (blue). Dashed lines correspond to results without postselection.}
    \label{fig:CodeCapacity}
\end{figure}

We evaluate the code capacity of the SOEO encoding. For this we simulate weight-one depolarizing noise on the four spins of the SOEO qubit [Eq.~\eqref{eq:error-on-single-qubit-gates}] and perform a perfect error-detection cycle, i.e., a perfect $XXXX$ and $ZZZZ$ readout \cite{aliferis2005quantumaccuracythresholdconcatenated, preskill1997faulttolerantquantumcomputation}. We run simulations for the input qubit states $\ket{\psi_{\rm in}} \in \{\ket{0}, \ket{1}, \ket{+}, \ket{-}, \ket{i}, \ket{-i}\}$ and postselect on all runs with trivial measurement outcomes. The acceptance rate is shown in Fig.~\ref{fig:CodeCapacity}a in red. We calculate the qubit error probability as $p_{\rm qubit} = 1 - \lvert\braket{\psi_{\rm in}}{\psi}\rvert^2$ with and without postselection and show the results in Fig.~\ref{fig:CodeCapacity}b as red solid and dashed lines. Without postselection, the results follow a linear trend $\propto p$. When postselecting on the trivial outcomes, linear error contributions are suppressed and the results follow a $\propto p^2$ behavior. 

Equivalently to the exchange-only qubit, SOEO qubit operations are usually constructed from several exchange pulses acting on more than a single spin within one encoded qubit \cite{Andrews_2019}. Thus, a single-spin depolarizing error model appears to be insufficient for the operations applied to SOEO to date. As a next step, we employ a four-spin depolarizing error channel to a perfectly initialized SOEO qubit followed by a noiseless round of quantum error detection. The depolarizing errors thus contain all possible errors that act on the four spins already at order $p_{\rm spin}$. We show the acceptance rate in Fig.~\ref{fig:CodeCapacity}a in blue. As expected, the acceptance rate is higher since weight-two $X$ and $Z$ errors, e.g. $XXII$, are not detected by the QED cycle. The qubit error probability is calculated with and without postselecting on trivial measurement outcomes and shown as solid and dashed lines in Fig.~\ref{fig:CodeCapacity}b. Since not all errors that occur with linear probability are detected, we obtain a linear scaling in the spin error probability for both curves. Although no quadratic suppression is achieved, we still find a reduction in the qubit error rate of almost one order of magnitude. Note that for the results without postselection, the single-spin error simulation performs worse, since all weight-one errors directly correspond to a qubit error. However, for the noise model with errors up to weight four, the errors $XXXX$ and $ZZZZ$ correspond to stabilizers and thus do not affect the qubit. 

\section{Fitting the qubit error probability}\label{App:FitErrorProbability}
In order to fit the idling error probability of the SOEO encoding, note that $m$ faults occur with probability $O(p_{\rm idle}^m)$. In the following analysis, we take into consideration up to four faults. The stabilizer readout can detect any weight-one fault, and some of the higher weight errors. The qubit error rate is then determined by the errors that cannot be detected and that are not equivalent to a stabilizer. The total number of single- and multi-weight spin errors is 255. After QED the remaining even-weight faults contain an even number of $X$, $Y$ and $Z$, respectively, as e.g. $XXII$, $ZZYY$, and $IZZI$. The weight-three faults that are not detected by the stabilizer readout are generated by weight-two $X$-operators and weight-two $Z$-operators. Thus all errors that contain exactly one $X$, one $Y$, and one $Z$ operator, i.e. $XYZI$, $XZIY$, or $IZYX$, are not detected by the stabilizer readout and lead to a qubit error.
We thus estimate the qubit error by counting all possible dangerous errors and error locations leading to 
\begin{align}
    p_{\rm qubit} = \frac{18}{54} p^2(1-p)^2 +  \frac{24}{108} p^3(1-p) +  \frac{21}{81} p^4. \label{Eq:FitErrorProbability}
\end{align}
The total number of non-detectable errors is 63 which corresponds to about 25\% of all possible errors. 
Note that the state overlap $\bra{\psi} E \ket{\psi}$ does not vanish for $E$ being a dangerous spin error. We thus expect the function in Eq.~\eqref{Eq:FitErrorProbability} to be an estimate for the qubit error rate. We use an ansatz of this form for fitting the memory experiment simulations in Fig.~\ref{fig:SOEO-QED-over-many-cycles}c and d.

\section{Scalable architectures} \label{App:shuttling}
\begin{figure}[ht]
    \centering
    \includegraphics[width=0.95\linewidth]{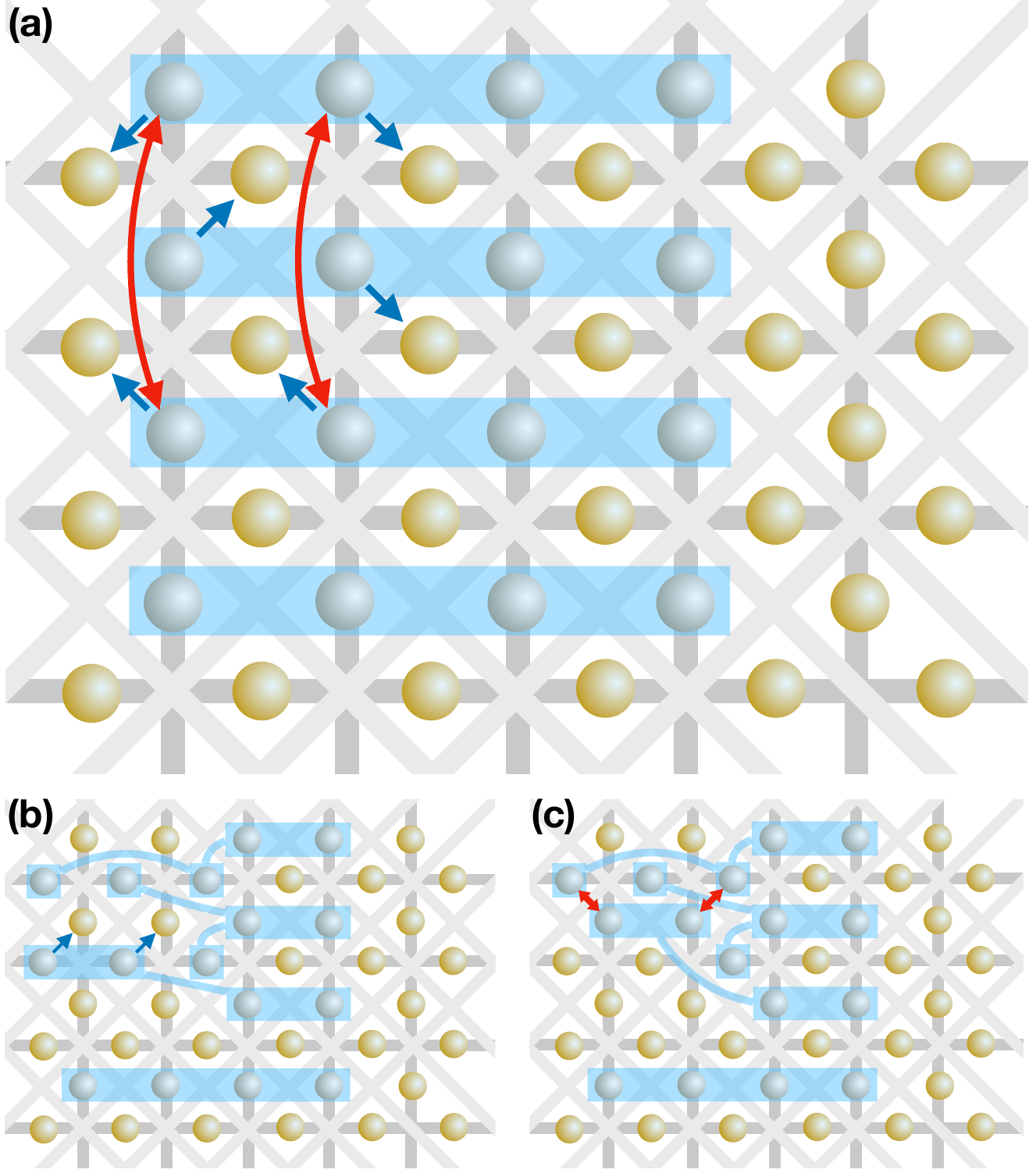}
    \caption{
    Schematic of a sparsely encoded SOEO qubit array that enables fault-tolerant routing of initially nonadjacent qubits into positions where the desired exchange interactions can be performed. The gray lines in the background indicate the electric control lines and each sphere indicates one electron spin. Blue boxes correspond to data SOEO qubits that are indirectly connected using the yellow auxiliary spins. Consider an example where the interactions indicated by the red arrows are desired for a two-qubit gate. For this, the spins of the SOEO qubits are shifted to the intermediate configuration shown in (b), as shown by the blue arrows, until they reach the final configuration in (c) where the desired exchange interactions can be applied.}
    \label{fig:SparseChip}
\end{figure}
To make use of the stabilizer readout proposed in this paper, we show an example of a layout on a chip that could be used to implement the surface code in Fig.~\ref{fig:Figure1}.  
In this figure, data SOEO qubits are shown in pink and auxiliary SOEO qubits in blue. Additional LD qubits can then perform error detection before coupling several SOEO qubits as required for the surface code. Single-spin operations would be performed in the LD qubit control zones. Per SOEO qubit, only the first auxiliary spin is shown. In general, one can introduce several additional spins. The flag spin can even remain in the interaction zone as it does not interact with the SOEO spins. After measuring trivial outcomes in the error detection step for each data and auxiliary qubit, the surface code stabilizers can be read out using CNOT gates between SOEO qubits. Note that inserting additional LD qubits and zones can optimize the protocol. We also point out that although CNOT gates between two SOEO qubits are not fault tolerant at this stage, the error detection scheme is still capable to detect some errors occurring during two-qubit gate sequences. 

We also introduce a sparsely encoded architecture in Fig.~\ref{fig:SparseChip}a where all-to-all connectivity is achieved using SWAP operations rather than shuttling. Here, the blue boxes represent one data SOEO qubit. To ensure that errors in two-qubit gates only affect at most two encoded SOEO qubits, additional spins (yellow) between data qubits (blue boxes) are introduced. These do not contain the logical information of the data qubits. As an example, we show how to enable the effective connectivity between the first-row and third-row data SOEO qubit indicated by the red arrows. For this, the spins are swapped in a way that the data spins never directly interact with the second-row SOEO qubit (Figs.~\ref{fig:SparseChip}b and c).  At first order in the error rate, this ensures that errors on the data qubits during the exchange pulses happen either only on the subspace of the desired interacting qubits while the second-row qubit is unaffected, or only on the second-row qubit while the interacting qubits are unaffected. The auxiliary spins potentially possess remaining errors and can be reset if necessary. Such protocols are important to ensure fault-tolerance for a higher-level quantum code. The efficiency of the layout can be improved and tailored to a specific quantum error-correcting code, enabling codes with higher connectivity also in a static gate setup. Stabilizer readout of the SOEO qubit could also be implemented using additional interaction zones placed on the chip, e.g., on the right side of the SOEO qubits.

%
%
\bibliography{bibliography.bib}

\end{document}